\documentclass[a4paper,11pt]{article}
\pdfoutput=1 % if your are submitting a pdflatex (i.e. if you have
\usepackage{jcappub} % for details on the use of the package, please
\usepackage[T1]{fontenc} % if needed
\usepackage{adjustbox}
\usepackage{float}
\usepackage{xcolor}

\newcommand{\be}{\begin{equation}}
\newcommand{\ee}{\end{equation}}
\newcommand{\ba}{\begin{eqnarray}}
\newcommand{\ea}{\end{eqnarray}}

\newcommand{\nus}{\nu_s}

\title{
Shedding Light on Dark Matter and Neutrino Interactions from Cosmology}

\author[a,1]{Arnab Paul }
\author[b,2]{, Arindam Chatterjee }
\author[c,3]{, Anish Ghoshal }
\author[a,d,4]{, Supratik Pal }

\affiliation[a]{Physics and Applied Mathematics Unit, Indian Statistical Institute, Kolkata-700108, India}
\affiliation[b]{Department of Physics,
School of Natural Sciences, Shiv Nadar University,\\ Gautam Budhha Nagar, Uttar Pradesh 201314, India}
\affiliation[c]{I. N. F. N. - Rome Tor Vergata, via della Ricerca Scientifica, I-00133 Rome, Italy}
\affiliation[d]{Technology Innovation Hub on Data Science, Big Data Analytics and Data Curation,\\ 
	Indian Statistical Institute, Kolkata-700108, India}

\emailAdd{arnabpaul9292@gmail.com}
\emailAdd{arindam.chatterjee@gmail.com}
\emailAdd{anishghoshal1@gmail.com}
\emailAdd{supratik@isical.ac.in}
\abstract{
In $\rm\Lambda$CDM cosmology, Dark Matter (DM) and neutrinos are assumed to be non-
interacting. However, it is possible to have scenarios, where DM-neutrino interaction
may be present, leading to scattering of  DM with neutrinos and annihilation of DM
into neutrinos. We investigate the viability of such scenarios in the light of cosmological
data by making use of the \texttt{Planck 2018} dataset (high-l TT+TE+EE, low-l TT, low-l EE) and constrain
these processes in the light of the same. We also discuss a viable particle DM model
where DM-neutrino interaction is present, and map the constraints obtained to the parameter
space of the model. 
}

\begin{document}

\maketitle

\newpage

\section{Introduction}
\label{intro}

The standard model of cosmology ($\Lambda$CDM) has been well established in the light 
of cosmological data, including precision observation of the Cosmic Microwave 
Background (CMB) anisotropy \cite{Aghanim:2018eyx}. In its simplest incarnation, the model 
incorporates a cosmological constant and a cold Dark Matter (CDM), which does not interact 
with the particle content of the Standard Model (SM) of particle physics. It has been 
estimated that the Dark Matter (DM) and the cosmological constant together constitutes approximately 
95\% of the energy budget of our Universe \cite{Aghanim:2018eyx}. 

On the other hand, while the standard model of particle physics does not include any suitable DM 
candidate, various well-motivated particle physics models incorporate particle DM 
candidates which also interacts with the SM particles non-gravitationally. In 
particular, Weakly Interacting Massive Particles (WIMP) DM candidates are postulated 
to be in the thermal equilibrium with the SM particles, thanks to their sizable 
interaction rate \cite{Bertone:2004pz}. The presence of interaction between DM and SM particles can 
lead to observable effects in the early Universe. For instance, the annihilation of 
the DM particles into the SM particles injects energy in the SM thermal bath, thereby 
affecting the CMB anisotropy \cite{Chluba_2010,Finkbeiner_2012,Galli:2013dna,Slatyer:2009yq}. Such interactions are also probed using direct \cite{Akerib:2016vxi,Cui:2017nnn,Aprile:2018dbl} 
and indirect searches \cite{Fermi-LAT:2016uux,Aguilar:2016kjl}. For the sub-GeV mass range, the most stringent limit 
on DM-baryon interaction strength (for s-wave annihilation process) has been 
obtained using the CMB data \cite{Aghanim:2018eyx}. Apart from DM annihilation scenarios, impact of 
DM-baryon scattering has also received some attention in the literature, both in 
the context of CMB \cite{Boddy:2018wzy,Boddy:2018kfv,Gluscevic:2017ywp} and 21cm cosmology \cite{Munoz:2015bca,Barkana:2018lgd,Fialkov:2018xre,Munoz:2018jwq,Munoz:2018pzp}.

Apart from DM-baryon interactions, several other scenarios have been considered 
in the literature constraining DM interactions with, for example, Dark Radiation 
(DR) \cite{Cyr-Racine:2015ihg}, in the light of cosmological data. DM-neutrino interactions \cite{Escudero:2015yka,Escudero:2018thh,Pandey:2018wvh} pose 
another interesting possibility. This is especially of interest since, such an 
interaction, if exists, would be difficult to probe (if not impossible, at least for certain DM candidates) using the terrestrial experiments\footnote{Interactions of neutrino with fuzzy DM has been considered, in the light of neutrino oscillation experiments, for example see ref. \cite{Brdar:2017kbt}. }. A particularly interesting scenario has been 
proposed in \cite{Berlin_2018,Berlin_2019}, where thermal production of a light (MeV) DM candidate has 
been considered, which rely on the interaction of DM with the (sterile) neutrino 
sector \cite{Berlin_2018,Berlin_2019}.

Along with DM-neutrino scattering, DM annihilation into rather highly energetic neutrinos have also been considered in 
the literature (see e.g. \cite{Cirelli:2005gh,Blennow:2007tw}), particularly in the context of indirect search for DM \cite{Arguelles:2019ouk,Arguelles:2019jfx,Frankiewicz:2015zma}. 
However, these probes only constrain the DM annihilation cross-section at the 
present epoch, and the limits are only stringent for rather heavy DM. 

Therefore, 
investigating such well-motivated possibilities in the light of cosmological data turn out to be a crucial project in recent times. This is particularly
important for two reasons : firstly, it enhances our understanding of the properties 
of DM; and secondly, including such additional well-motivated 
parameters to the simplest cosmological model also helps in establishing the 
robustness of the (other) estimated parameters, especially in case of possible 
degeneracies.

In this article, we consider a scenario where DM interacts with (sterile) neutrinos. 
In particular, we assume that the (sterile) neutrinos are light $(\mathcal{O}$(1 eV))
and will only decay, if at all, into invisible decay channels which do not inject 
energy to the baryon-photon plasma in the early Universe. Further, we assume that the 
interaction is mediated by a light (pseudo-scalar) boson, which is also essential 
for the viability of such light sterile neutrinos \cite{Babu:1991at,Enqvist:1992ux,Hannestad:2013ana,Archidiacono:2014nda,Archidiacono:2015oma,Archidiacono:2016kkh}. In our scenario, this mediator 
also gives rise to Sommerfeld enhanced annihilation of the DM-particles \cite{Tulin:2013teo,Hisano:2004ds,Cirelli:2007xd,ArkaniHamed:2008qn,Feng:2009hw,Bringmann:2016din,Diamanti:2013bia,Liu:2016cnk}. In presence 
of the drag term (as a consequence of DM- (sterile) neutrino scattering) and 
(Sommerfeld enhanced) annihilation term, we consider the evolution of perturbations via the Boltzmann 
equations in detail to estimate the impact on CMB temperature anisotropy, thereby 
constraining such interactions in the light of CMB. There have been some previous
studies in the literature \cite{Mangano_2006,Wilkinson:2014ksa,DiValentino:2017oaw,Olivares_Del_Campo_2018,Stadler:2019dii,Ghosh:2019tab,Mosbech:2020ahp} and \cite{Armendariz_Picon_2012,Bringmann_2018,Cui:2018imi} considering a similar scenario. As we will describe 
in detail in sec \ref{sec_pert}, we improve on the previous studies by incorporating the effect 
of both the drag term, as well as the time varying annihilation term for the DM particle, also by constraining the parameters by latest cosmological data till date.

The paper is organized as follows: in section \ref{sec_pert} we discuss the cosmological scalar perturbation equations when DM-neutrino (DM-$\nu$) scattering and DM annihilation is present, in section \ref{analysis} the change in the observables (namely CMB Temperature-Temperature (TT) and Matter Power Spectrum) is discussed and constraints of the parameters in the light of the \texttt{Planck 2018} dataset (high-l TT+TE+EE, low-l TT, low-l EE)
are presented in section \ref{results}. In section \ref{model}, a model where these effects come naturally is discussed and the constraints of section \ref{results} to the parameter space of the model is mapped.

\medskip

%%%%%%%%%%%%%%%%%%%%%%%%%%%%%%%%%%%%%%%%%%%%%%%%%%%%%%

\section{Evolution of scalar perturbations}
\label{sec_pert}

In order to study the evolution of scalar perturbations in presence of DM-$\nu$ scattering or DM annihilation, as the case may be,
one  needs to extend the 6-parameter vanilla $\Lambda $CDM model so as to accommodate the effects of scattering or annihilation. 
For any DM model having sufficient DM-$\nu$ interaction, denoting DM by $\Psi$, 
%sterile neutrino by $\nu_s$, and the mediator between the dark sector \& $\nu$ by $\phi$, 
we expect the following 
modifications to the standard $\Lambda $CDM paradigm:

\begin{itemize}
    \item The co-moving densities of different species change throughout the evolution history we are concerned about.
    \item The above-mentioned species scatter with each other and hence do not behave as independent species, especially the evolution of scalar perturbations of DM and neutrinos may deviate from the standard $\rm\Lambda$CDM case.
    \item The DM particles, depending on the model, %i.e. if the number density of DM is substantial and the mass of $\phi$ is small,
     may result in a Sommerfeld enhanced annihilation at late times.
     \item Both DM and neutrino species may have self scattering. 
\end{itemize}

We are mainly interested in the effects of DM-$\nu$ scattering and enhanced annihilation of DM particles at late time on the scalar perturbation evolution.
In the fluid-like approach, only density contrast and velocity divergence perturbation of DM are sufficient to be considered without higher-order multipoles \cite{Ma_1995}. We neglect self scattering of DM or neutrinos in this work. 

In this work, to speed up computation, we assume the neutrinos to be relativisitc species throughout the history of the Universe. It has been shown recently in ref. \cite{Mosbech:2020ahp} that this assumption does not change the results considerably, so this is indeed a viable assumption one can use.

\subsection{DM-neutrino scattering}

Scattering between two species results in a drag term in the perturbed Euler equations for the corresponding species. In Synchronous gauge, we get the velocity divergence $\theta_{\rm DM}$ of DM to have a possible zero solution throughout the evolution. This enables us to study the scenario with a system having one less variable, namely $\theta_{\rm DM}$. But, once DM scatters with some other species, for example neutrinos, we do not have the zero solution of $\theta_{\rm DM}$ due to the extra drag term. Although, it may still be possible to set the Synchronous gauge frame to simplify the set of equations,  in this work we do not explore that path, rather we choose Newtonian gauge for our analysis.
In Newtonian gauge, assuming flat Universe, the density and velocity perturbation equations in presence of scattering between DM and neutrino, is given by \cite{Stadler:2019dii}:
\begin{eqnarray}
\label{eqn_DMnu}
\dot \delta_{\rm{DM}} &=& -\theta_{\rm{DM}} + 3\dot \phi,\nonumber\\
\dot \theta_{\rm{DM}}
& = &   k ^2\psi - {\cal H} \theta_{\rm{DM}} - S^{- 1} \dot \mu (\theta_{\rm{DM}} - \theta_\nu)~,\nonumber \\
\dot \delta_{\rm{\nu}} &=& -\frac{4}{3}\theta_{\rm{DM}} + 4\dot \phi,\nonumber\\
\dot \theta_\nu
& = &   k^2 \psi + k^2 \left(\frac{1}{4} \delta_\nu - \sigma_\nu\right) - \dot \mu (\theta_\nu - \theta_{\rm DM})~ 
\end{eqnarray}
where the derivatives are with respect to conformal time. Here,
$\delta_{\rm{DM}}$ and $\delta_\nu$ are, respectively,  the DM and neutrino density fluctuations, $\theta_{\rm{DM}}$ and $\theta_\nu$ are the corresponding DM and neutrino velocity divergences, $k$ is the co-moving wavenumber, $\psi$ is the gravitational potential, $\sigma_\nu$ is the neutrino anisotropic stress potential, and ${\cal H}=(\dot a / a)$ is the conformal Hubble parameter.

Further, $\dot{\mu} \equiv a\hspace{0.3ex}\sigma_{\Psi-\nu}\hspace{0.3ex}c\hspace{0.3ex}n_{\rm{DM}}$ is the DM-$\nu$ scattering rate, where $\sigma_{\Psi-\nu}$ is the elastic scattering cross section, $n_{\rm{DM}} = \rho_{\rm{DM}} / M_{\Psi}$ is the DM number density, $\rho_{\rm{DM}}$ is the DM energy density and $M_{\Psi}$ is the DM mass. As usual,  $S$ is given by $(3/4)(\rho_{\rm{DM}}/\rho_\nu)$.  The DM-$\nu$ scattering also affects the hierarchy of Boltzmann equations for neutrino species. 

%We note that, although in general 
%the scattering cross section between DM and neutrinos, $\sigma_{\Psi-\nu}$, may be velocity dependent, %$v^{\alpha}$ (where $\alpha$ may be positive or negative for Sommerfeld enhanced cross-section). This will depend on the specific particle physics model being considered. For our study,
Although we consider $\sigma_{\Psi-\nu}$ to be constant in this study, 
 in general 
$\sigma_{\Psi-\nu}$ may be velocity dependent \cite{Wilkinson:2014ksa}.
The effect of DM-$\nu$ scattering is quantified by a dimensionless quantity:
\begin{equation}
\label{u_nudm}
u \equiv \frac{\sigma_{\Psi-\nu}}{\sigma_{\rm Th}}  \frac{100~\rm{GeV}}{M_{\Psi}} ~,
\end{equation}
where $\sigma_{\rm Th}$ is the Thomson cross section.

\subsection{DM annihilation and Sommerfeld enhancement}

As soon as the DM annihilation rate becomes lower than the Hubble expansion rate, the resulting number density of DM particles per unit co-moving volume thereafter becomes constant. However if there is some enhancement process \cite{Tulin:2013teo,Hisano:2004ds,Cirelli:2007xd,ArkaniHamed:2008qn,Feng:2009hw,Bringmann:2016din,Diamanti:2013bia,Liu:2016cnk}  like non-perturbative effects, DM annihilation via light (pseudo) scalar particle may get boosted. Particularly, at late times, Sommerfeld enhancement may provide palpable effects in the CMB power spectra \cite{Armendariz_Picon_2012,Bringmann_2018}. 

Incorporating Sommerfeld enhancement in the annihilation of DM particles results in a modified density and velocity perturbation equations
for DM and neutrinos. In the Newtonian gauge they read,

\begin{eqnarray}
\label{eqn_DMnu_DManni}
\dot \delta_{\rm{DM}} &=& -\theta_{\rm{DM}} + 3\dot \phi-\frac{\delta\langle\sigma v\rangle}{M_{\Psi}}\rho_{\rm DM}a-\frac{\langle\sigma v\rangle}{M_{\Psi}}\rho_{\rm DM}\delta_{\rm DM}a-\frac{\langle\sigma v\rangle}{M_{\Psi}}\rho_{\rm DM}a\psi ~,\nonumber\\
\dot \theta_{\rm{DM}}
& = &   k ^2\psi - {\cal H} \theta_{\rm{DM}} - S^{- 1} \dot \mu (\theta_{\rm{DM}} - \theta_\nu)+2\frac{\langle\sigma v\rangle}{M_{\Psi}}\rho_{\rm DM}\theta_{\rm DM}a~,\nonumber \\
\dot \delta_{\rm{\nu}} &=& -\frac{4}{3}\theta_{\rm{DM}} + 4\dot \phi+\frac{\delta\langle\sigma v\rangle}{M_{\Psi}}\frac{\rho_{\rm DM}^2}{\rho_{\nu}}a+\frac{\langle\sigma v\rangle}{M_{\Psi}}\frac{\rho_{\rm DM}^2}{\rho_{\nu}}\left(2\delta_{\rm DM}-\delta_\nu \right)a+\frac{\langle\sigma v\rangle}{M_{\Psi}}\frac{\rho_{\rm DM}^2}{\rho_{\nu}}a\psi ,\nonumber\\
\dot \theta_\nu
& = &   k^2 \psi + k^2 \left(\frac{1}{4} \delta_\nu - \sigma_\nu\right) - \dot \mu (\theta_\nu - \theta_{\rm DM})-a\frac{\langle\sigma v\rangle}{M_{\Psi}}\frac{\rho_{\rm DM}^2}{\rho_{\nu}}\left(\frac{3}{4}\theta_{\rm DM}+\theta_\nu \right)~ 
\end{eqnarray}
where $\langle\sigma v\rangle$ is the thermally averaged annihilation cross-section of DM particles and $M_{\Psi}$ is the mass of the DM particle. We paramertise $\langle\sigma v\rangle$ as
\be
\frac{\langle\sigma v\rangle}{M_{\Psi}}\frac{100~{}\rm GeV}{\langle\sigma v\rangle_{\rm w}}\sim\Gamma a~,
\ee
where $\langle\sigma v\rangle_{\rm w}=3\times10^{-26}~\rm cm^3 s^{-1}$, the proportionality factor of $a$ comes due to Sommerfeld enhancement \cite{Tulin:2013teo,Hisano:2004ds,Cirelli:2007xd,ArkaniHamed:2008qn,Feng:2009hw,Bringmann:2016din,Diamanti:2013bia,Liu:2016cnk}, $\Gamma$ being a dimensionless constant \footnote{The values fo $\Gamma$ which we mention throughout this work, has an additional constant factor attached to it, the relation with this additional factor is given by, $$3\times10^{-12}\frac{\langle\sigma v\rangle}{M_{\Psi}}\frac{100~{}\rm GeV}{\langle\sigma v\rangle_{\rm w}}=\Gamma a.$$  The constant factor $3\times10^{-12}$ in the front represents some rescaling factors in the code \texttt{CLASS}.}. %Similar to the DM-$\nu$ interaction rate, $\Gamma$ may also be velocity dependent, for our preliminary study we have assumed it to be constant.

Further, although neutrinos may self-scatter through the scalar mediator $\phi$, we consider the effects to be negligible as this rate is proportional to $g_s^4$ and $g_s$ is chosen to be very small, $g_s\gtrsim\mathcal{O}(10^{-5})$ \cite{Archidiacono:2016kkh}. For the case of DM-DM scattering, DM particles can only feel non-zero force if it is surrounded asymmetrically by others. Hence the DM scattering effect comes into play only if we consider higher multipoles (higher than 2) of the perturbation. We neglect this effect too in this present analysis.

What turns out from the above discussion is that, effectively we need to introduce two new parameters, $u$ and $\Gamma$, on top of the 6 parameter vanilla $\Lambda$CDM scenario.  
As we will discuss in  section \ref{results}, we implement necessary modifications to publicly available code \texttt{CLASS}  \cite{Blas_2011} 
in order to accommodate
these effects in the perturbation equations for individual species considering DM-$\nu$ scattering and DM annihilation in a 6+2 parameter description. We further use the MCMC code \texttt{MontePython} \cite{Brinckmann:2018cvx} to estimate the parameters under consideration with their posterior distributions
and to analyse our results vis-a-vis $\Lambda$CDM.

\medskip

%%%%%%%%%%%%%%%%%%%%%%%%%%%%%%%%%%%%%%%%%%%%%%%%%%%%%%%%%%

\section{Analysis of CMB TT and matter power spectra}
\label{analysis}

\subsection{Effect of DM-neutrino scattering}

The CMB power spectrum (PS) primarily depends on the sum of temperature fluctuations of photons which is proportional to their density fluctuations and metric perturbation $\psi$ at the last scattering surface, Doppler effects due to velocity perturbation of baryons at that time and integrated Sachs-Wolfe effect due to the evolution of the metric perturbations throughout the path of the photon from last scattering surface to us.

The significance of scattering between the DM and the neutrino species, as also discussed elaborately in \cite{Wilkinson:2014ksa}, is manifest in the following ways in CMB PS.
\begin{itemize}
    \item Presence of a DM-$\nu$ scattering results in clustering (in contrast to free-streaming) of the neutrinos during radiation domination. This clustering gives rise to deeper gravitational potential fluctuations, affecting the photon-baryon fluid oscillations with a gravitational boost effect. This effect enhances all the peaks barring the first one.
    \item Due to the DM-$\nu$ scattering, the combined fluid attains a sound speed, which is smaller than that of the baryon-photon fluid as non-relativistic matter fraction is higher in the first fluid. This in turn drags back the sound waves of the photon-baryon fluid, letting them move to a smaller distance than for the standard scenario. So, the peaks get shifted to larger $l$ values. 
    \item The contrast between even and odd peaks of CMB PS occurs due to shift of temperature fluctuation oscillations by metric perturbation. Having DM-$\nu$ scattering during recombination suppresses the metric perturbation, hence decreasing the contrast between even and odd peaks.
    \item  Metric  fluctuations evolve with  time  as  long  as  DM  stays coupled to neutrinos efficiently. As stated before, CMB PS gets contribution from variation of metric  fluctuations along the photon path known as integrated Sachs-Wolfe effect. Variation of this kind just after recombination (after the mode corresponding to the first peak has entered the horizon) thus results in further enhancement of the first peak.
    \item The features of the tail of CMB PS is highly suppressed due to diffusion damping, so no visible difference due to non-standard scatterings is present at those modes. On the other hand, small $l$ modes are mainly dependent on the initial PS and late time evolution of the Universe (as these modes enter the horizon at late time and also the late integrated Sachs-Wolfe effect due to metric fluctuation variation during Matter-Dark Energy equality contributes to these modes). As these are not much affected by the DM-$\nu$ scattering, changing the scattering strength does not affect low $l$ modes.
\end{itemize}

\begin{figure}[H]
	\centering
	\includegraphics[width=0.48\textwidth]{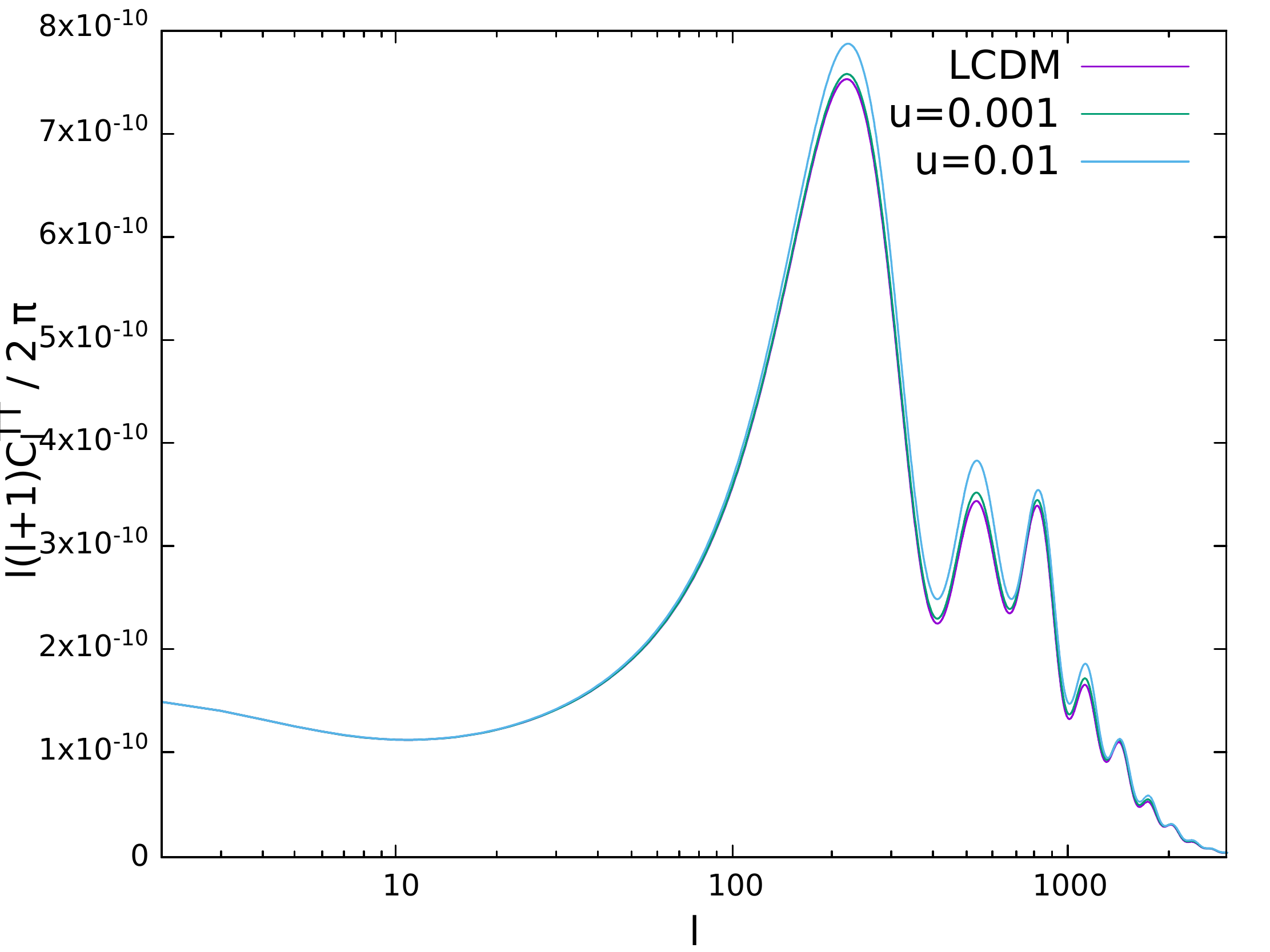}
    \includegraphics[width=0.48\textwidth]{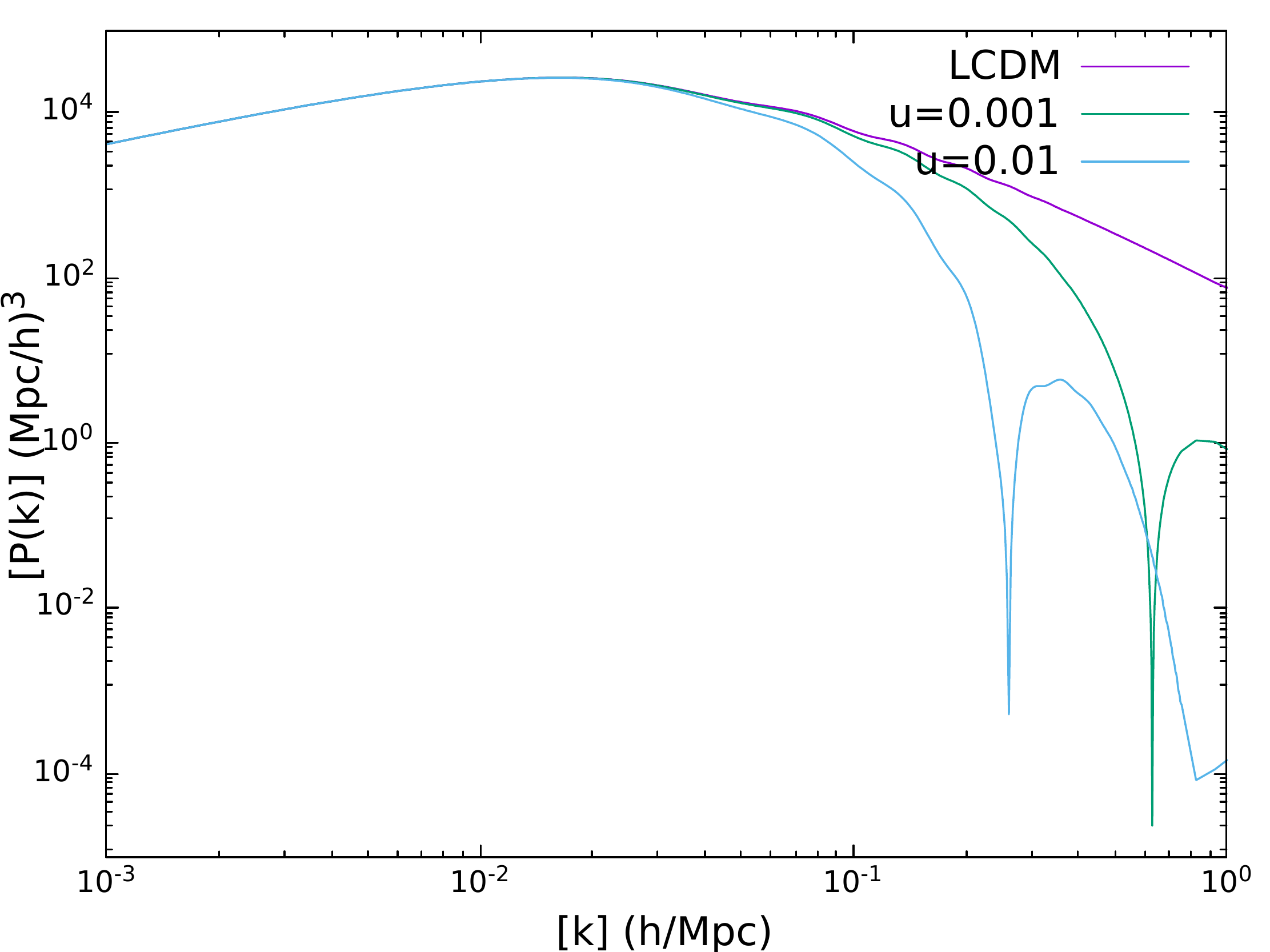}
	\caption{\it The effect of DM-$\nu$ scattering on the $TT$ power spectrum (left panel) and on matter power spectrum (right panel). We keep $\Gamma=0$ for these plots.}
	\label{fig:clpk1}
\end{figure}

The qualitative nature of the effects of DM-$\nu$ scattering on CMB TT PS  has been shown in fig. \ref{fig:clpk1} : left panel. The change in height of the peaks compared to the vanilla 
$\Lambda$CDM case
are clearly visible, as expected from the above discussions.

In fig. \ref{fig:clpk1} : right panel, we show the plot of matter PS due to different scattering strength $u$ along with the one expected from vanilla 
$\Lambda$CDM model.
As observed in the matter power spectrum P(k) plot, an increase in  $u$ results in a suppression of  structure formation at small length scales (large k). This happens because due to scattering with relativistic neutrinos, the dark matter particles are dragged out of the potential wells and can not clump as much as it would have, if the scattering was absent. We also get an oscillatory behaviour at large k values, similar to Dark Acoustic Oscillations \cite{Cyr-Racine:2013fsa}.

\subsection{Effect of DM annihilation}
%\label{}

\begin{figure}
	\centering
	\includegraphics[width=0.48\textwidth]{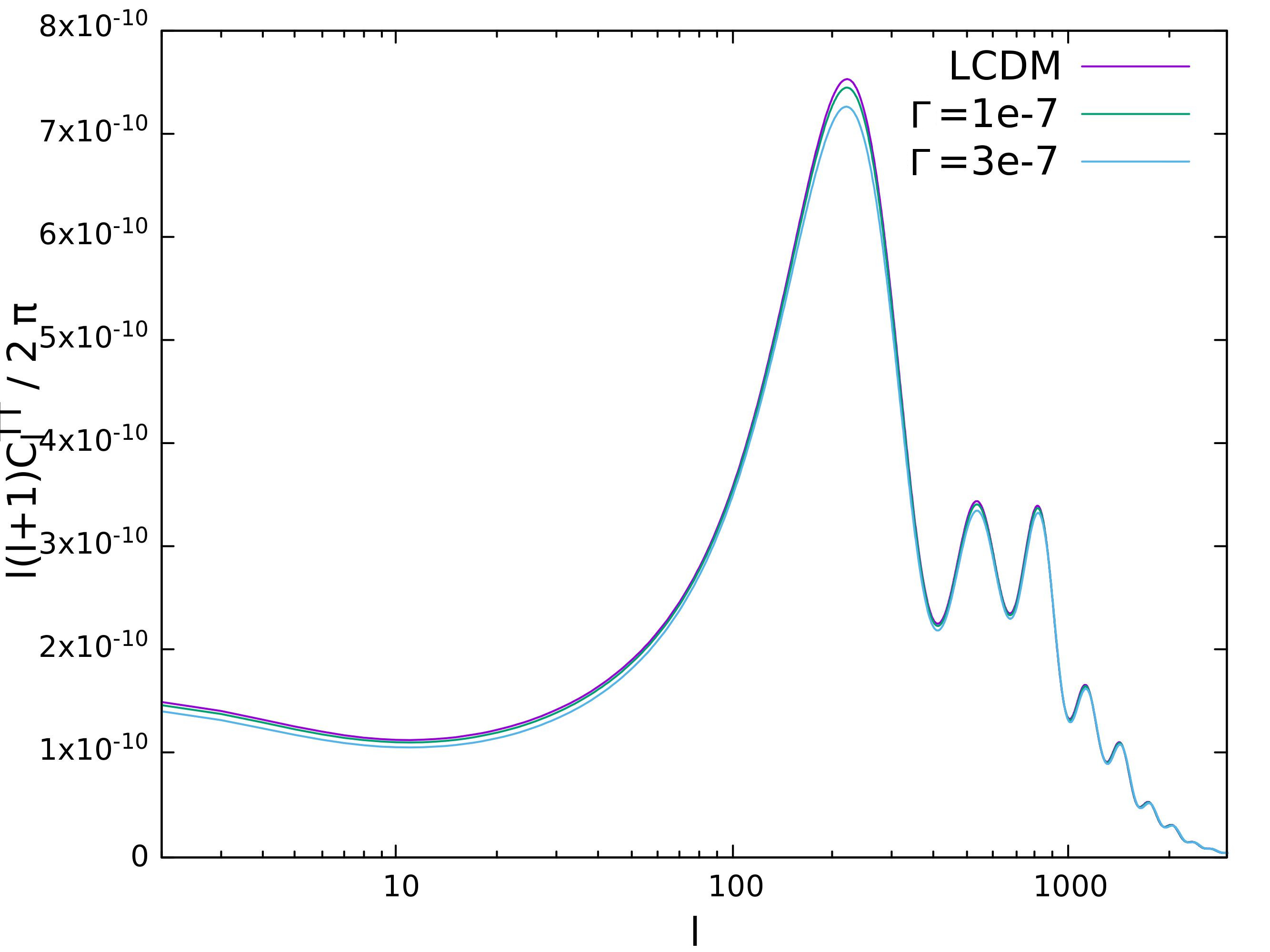}
    \includegraphics[width=0.48\textwidth]{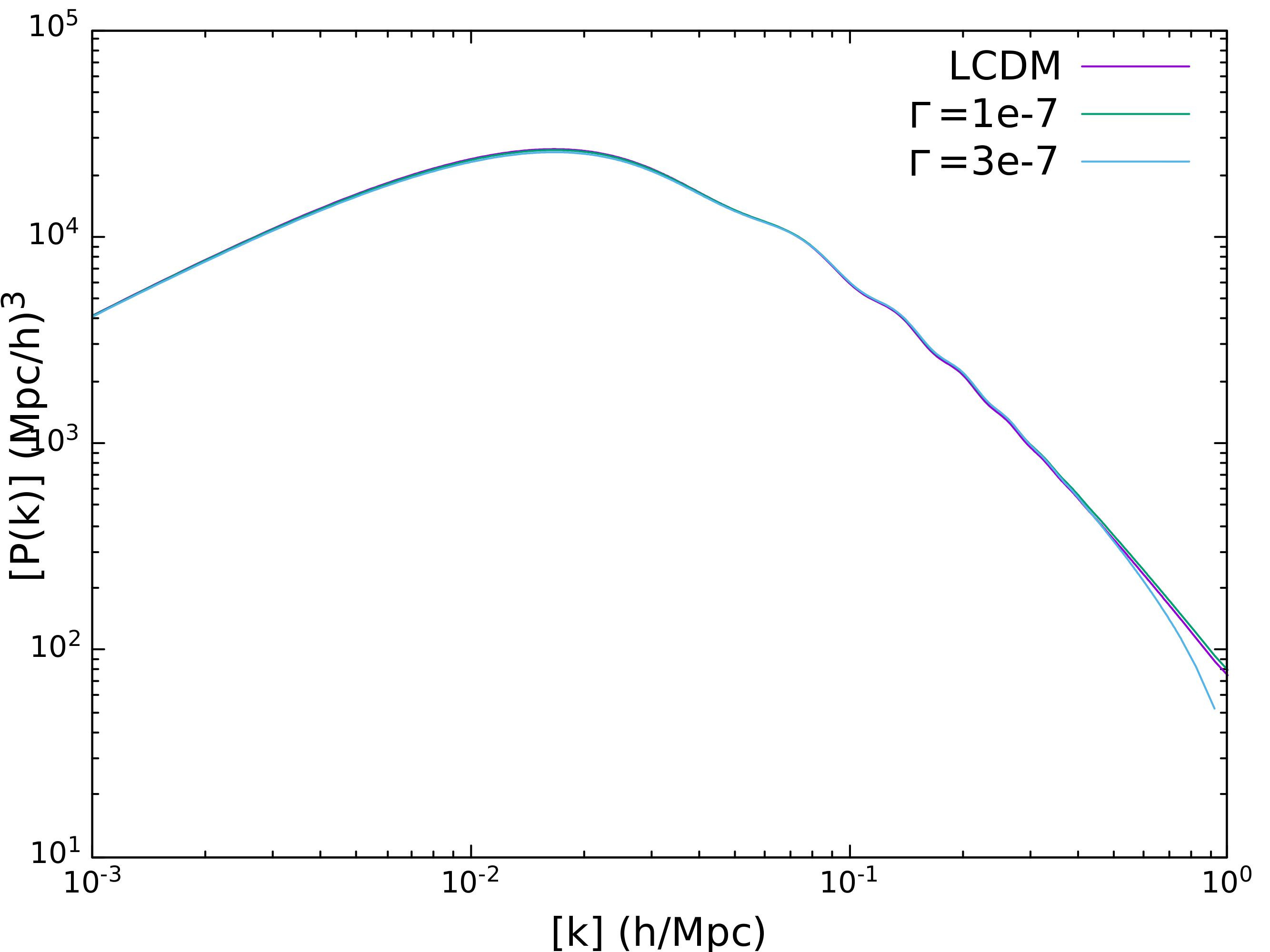}
	\caption{\it The effect of DM annihilation on the $TT$ power spectrum (left panel) and on matter power spectrum (right panel). We keep $u=0$ for these plots. }
	\label{fig:clpk2}
\end{figure}

Effects of
DM annihilation on CMB TT (left panel) and matter PS (right panel) have been  shown in fig. \ref{fig:clpk2}. One can see that
 the height of the peaks of CMB PS is lowered when the annihilation channels of the DM particles are open (left panel), resulting in less power in TT correlations. This is primarily because of the fact that  annihilation of DM into relativistic species decreases the DM density perturbations resulting in shallower gravitational potential fluctuations. Therefore the baryon acoustic oscillations (BAO) take place in a potential well which would have been deeper if DM annihilation was not present, so the gravitational boost effect is smaller in the scenario where DM annihilates. This effect reduces the heights of the peaks, as clearly visible in fig. \ref{fig:clpk2}. This also decreases the contrast between even and odd peaks, similar to the scenario in DM-$\nu$ scattering.

Similar to the DM-$\nu$ scattering we observe that in the P(k) plot (fig.  \ref{fig:clpk2} : right panel), increment of $\Gamma$ suppresses structure formation at small length scales (large k). The non-relativistic nature of CDM make the density perturbations grow, whereas if a fraction of CDM particles annihilate into relativistic particles, they free-stream from the higher density region, resulting in a dip in the power of density fluctuation. 

%%%%%%%%%%%%%%%%%%%%%%%%%%%%%%%%%%%%%%%%%%%%%%%%%%%%%%%%

\section{Methodology, statistical results and cosmological observables}
\label{results}

As mentioned earlier, in order to study the evolution of scalar perturbations in presence of DM-$\nu$ scattering or DM annihilation, as the case may be,
one  needs to extend the 6-parameter vanilla $\Lambda $CDM scenario so as to accommodate the additional degrees of freedom.  
Below we discuss the methodology used to incorporate the changes and the results obtained therefrom.

In order to accommodate the changes due to DM-$\nu$ scattering and DM annihilation to vanilla $\Lambda $CDM model in the numerical analysis, 
we have made use of  the  publicly available modified version \cite{Stadler:2019dii} of the code \texttt{CLASS} \cite{Blas_2011} and used the MCMC code \texttt{MontePython} \cite{Brinckmann:2018cvx}.
We implemented necessary modifications in \texttt{CLASS} to accommodate the modified perturbation equations of section \ref{sec_pert}. This will constrain the DM-$\nu$ scattering strength $u$ and DM annihilation rate $\Gamma$ from cosmological observables.

We have taken the \texttt{Planck 2018} dataset (high-l TT+TE+EE, low-l TT, low-l EE) under consideration in order  to get the posterior distributions of the parameters under consideration. 
As already stated, we first deal with 
a 6+2 parameter model of cosmology that accounts for the phenomena, namely,  DM-$\nu$ scattering and DM annihilation. 
The set of cosmological parameters under consideration are:
$\lbrace \omega_{\rm b},~\omega_{\rm cdm},~\theta_{\rm s},~A_{\rm s},~n_{\rm s},~\tau_{\rm reio},~u,~\Gamma \rbrace$.
% presented in figures \ref{1d_6ug} \& \ref{triangle_6ug}. The statistical results are given in the table \ref{table1}.

%\subsection{6+2 parameter distributions}

\subsection{Likelihood analysis \& parameter degeneracies in 6+2 parameter model }

\begin{figure}
	\centering
	\includegraphics[width=1.\textwidth]{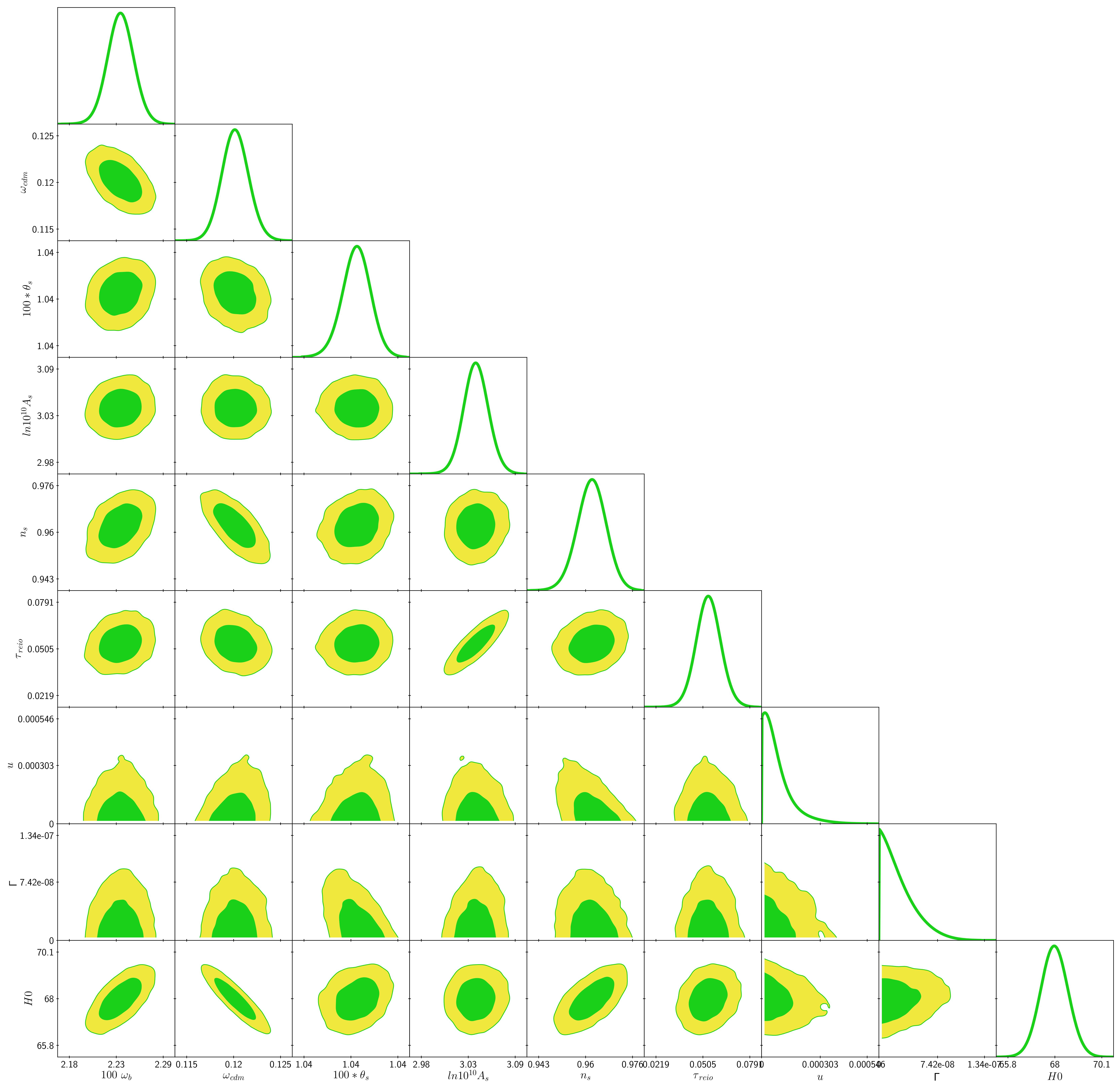}
	\caption{\it 2-d posterior distributions of 6+2 parameter model with parameters $\lbrace \omega_{\rm b},~\omega_{\rm cdm},~\theta_{\rm s},~A_{\rm s},~n_{\rm s},~\tau_{\rm reio}, ~u, ~\Gamma \rbrace$ using \texttt{Planck 2018} dataset (high-l TT+TE+EE, low-l TT, low-l EE).}
	\label{triangle_6ug}
\end{figure}

As discussed earlier, we have used the \texttt{Planck 2018} dataset (high-l TT+TE+EE, low-l TT, low-l EE) to get the posterior distributions of the parameters for 6+2 parameter model  $\lbrace \omega_{\rm b},~\omega_{\rm cdm},~\theta_{\rm s},~A_{\rm s},~n_{\rm s},~\tau_{\rm reio},~u,~\Gamma \rbrace$. The priors of the newly introduced parameters for the Monte-Carlo analysis are flat, with lower and upper limit as follows:
\begin{eqnarray}
u &=& [0,10^{-3}]\nonumber\\
\Gamma &=& [0,5\times10^{-7}]\nonumber
\end{eqnarray}
We will see later that the upper limits of the priors are well above the 95\% upper bound of the posterior distribution, hence justifying our choice of priors for quick convergence of the MCMC chains. In the light of the discussions of the effect of DM-$\nu$ scattering and DM annihilation in the previous section and noting the fact that both the effects tend to suppress matter power spectrum at small scales, we expect to get only upper bounds of the new parameters. It is clearly evident from fig. \ref{triangle_6ug} that we only find upper bounds on the new positive parameters $u$ and $\Gamma$. 

Table \ref{tab:table1} 
summarizes the major statistical results of our analysis. They include best fit values as well as 
mean values with  $1-\sigma$ error for the parameters of the cosmological model, it should be noted that for the non-standard parameters $u$ and $\Gamma$, we only mention $1-\sigma$ and $2-\sigma$ upper bounds, as the maxima of the posterior distribution is close to $0$: the theoretical lower bound. This reiterates that $\rm\Lambda$CDM  is still the simplest model under consideration and we can at best put some upper bounds on the interaction strengths as far as the present cosmological data is concerned. This is true at least from the perspective of the cosmological observables we have taken under consideration.

\begin{table}[H] 
\begin{tabular}{|l|c|c|c|c|} 
	\hline 
	Parameter & best-fit & mean$\pm\sigma$ & 95\% lower & 95\% upper \\ \hline 
	$100~\omega{}_{b }$ &$2.239$ & $2.239_{-0.016}^{+0.015}$ & $2.21$ & $2.269$ \\ 
	$\omega{}_{cdm }$ &$0.1201$ & $0.1204_{-0.0015}^{+0.0014}$ & $0.1176$ & $0.1233$ \\ 
	$100*\theta{}_{s }$ &$1.042$ & $1.042_{-0.00034}^{+0.00038}$ & $1.041$ & $1.042$ \\ 
	$ln10^{10}A_{s }$ &$3.04$ & $3.045_{-0.017}^{+0.016}$ & $3.013$ & $3.077$ \\ 
	$n_{s }$ &$0.9638$ & $0.9618_{-0.005}^{+0.0055}$ & $0.9514$ & $0.9721$ \\ 
	$\tau{}_{reio }$ &$0.05354$ & $0.05373_{-0.008}^{+0.0074}$ & $0.03856$ & $0.06937$ \\ 
	$u$ &$-$ & $.0001003~(\rm 1-\sigma ~upper)$ & $-$ & $0.0002373$ \\ 
	$\Gamma$ &$-$ & $3.204\times10^{-8}~(\rm 1-\sigma ~upper)$ & $-$ & $6.821\times10^{-8}$ \\ 
	$H0$ &$67.96$ & $67.95_{-0.66}^{+0.62}$ & $66.7$ & $69.22$ \\ 
	\hline 
\end{tabular} \\\caption{\it Statistical results of 6+2 parameter model with parameters $\lbrace \omega_{\rm b},~\omega_{\rm cdm},~\theta_{\rm s},~A_{\rm s},~n_{\rm s},~\tau_{\rm reio}, ~u, ~\Gamma \rbrace$ using \texttt{Planck 2018} dataset (high-l TT+TE+EE, low-l TT, low-l EE).}
\label{tab:table1} 
\end{table}

%THIS TABLE BELOW HAS MEANS ALSO
%\begin{table}[H] \label{table1}
%\begin{tabular}{|l|c|c|c|c|} 
%\hline 
%Parameter & best-fit & mean$\pm\sigma$ & 95\% lower & 95\% upper \\ \hline 
%$100~\omega_{b }$ &$2.249$ & $2.244_{-0.022}^{+0.022}$ & $2.2$ & $2.287$ \\ 
%$\omega_{cdm }$ &$0.1183$ & $0.1184_{-0.0014}^{+0.0014}$ & $0.1156$ & $0.1212$ \\ 
%$100*\theta_{s }$ &$1.042$ & $1.042_{-0.0005}^{+0.00056}$ & $1.041$ & $1.043$ \\ 
%$ln10^{10}A_{s }$ &$3.193$ & $3.183_{-0.052}^{+0.059}$ & $3.07$ & $3.29$ \\ 
%$n_{s }$ &$0.97$ & $0.9661_{-0.0061}^{+0.0063}$ & $0.9539$ & $0.9782$ \\ 
%$\tau_{reio }$ &$0.13$ & $0.1251_{-0.027}^{+0.03}$ & $0.06748$ & $0.18$ \\ 
%$u$ &$9.445e-06$ & $0.0001025_{-0.0001}^{+2e-05}$ & $1.941e-09$ & $0.0002948$ \\ 
%$\Gamma$ &$2.268e-08$ & $5.998e-08_{-6e-08}^{+1.7e-08}$ & $2.087e-12$ & $1.501e-07$ \\ 
%$H_0$ &$68.81$ & $68.97_{-0.69}^{+0.67}$ & $67.6$ & $70.33$ \\ 
%\hline 
% \end{tabular} \\\caption{\it Statistical results of 6+2 parameter model with parameters $\lbrace \omega_b,~\omega_{cdm},~\theta_s,~A_s,~n_s,~\tau, ~u, ~\Gamma \rbrace$ using \texttt{Planck 2018 high-l TT, BOSS-BAO 2014} data-sets.}
%\end{table}
%
%The 1-d and 2-d posterior distributions  for 6+2 parameter model  $\lbrace \omega_b,~\omega_{cdm},~\theta_s,~A_s,~n_s,~\tau,~u,~\Gamma \rbrace$ have been  presented in figures \ref{1d_6ug} \& \ref{triangle_6ug} respectively. 

\begin{figure}
	\centering
	\includegraphics[height=10cm,width=10cm]{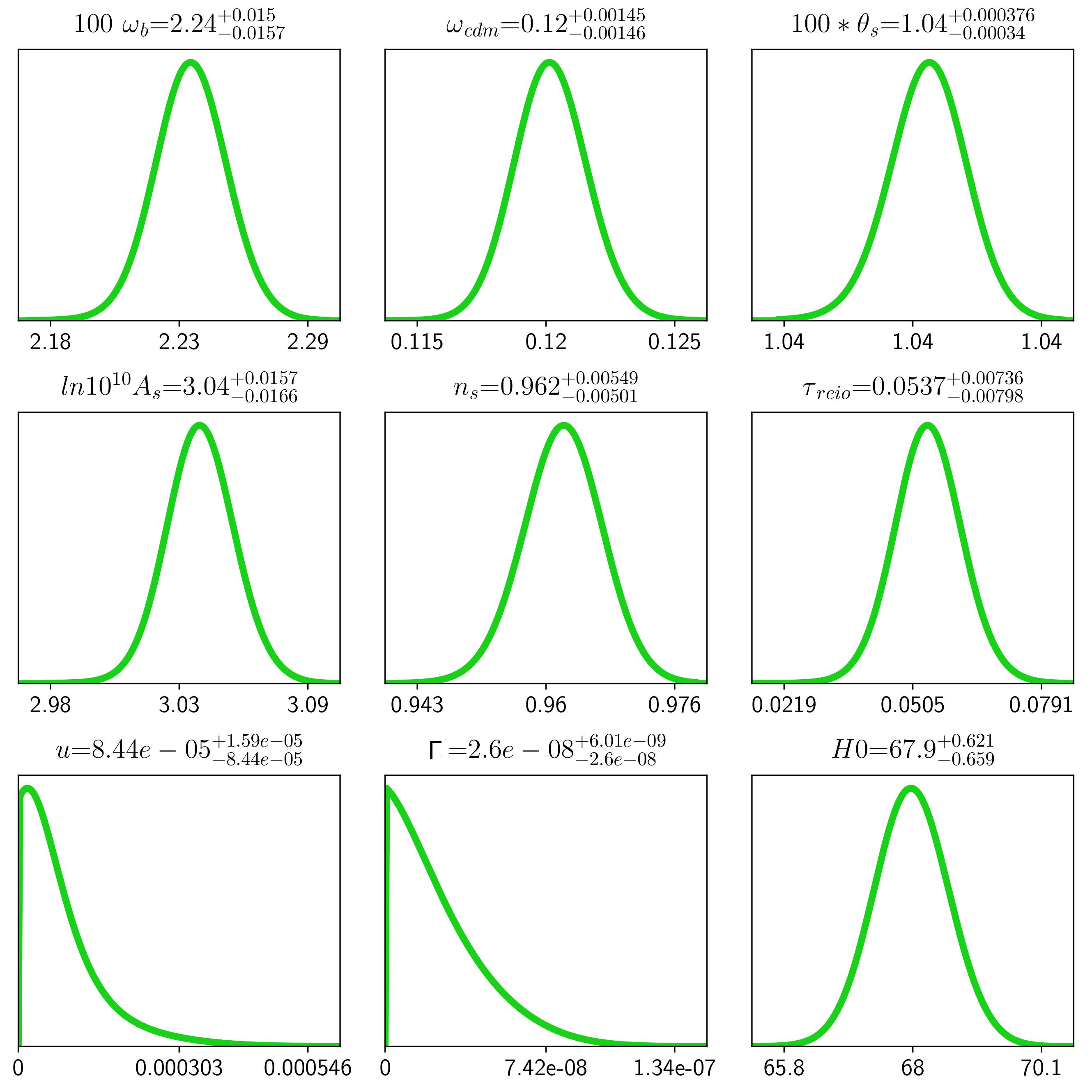}
	\caption{\it 1-d posterior distributions of 6+2 parameter model with parameters $\lbrace
		\omega_{\rm b},~\omega_{\rm cdm},~\theta_{\rm s},~A_{\rm s},~n_{\rm s},~\tau_{\rm reio}, ~u, ~\Gamma \rbrace$ using \texttt{Planck 2018} dataset (high-l TT+TE+EE, low-l TT, low-l EE).}
	\label{1d_6ug}
\end{figure}

%\noindent\textbf{Correlation with cosmological parameters}

%\noindent 
Let us now try to understand the correlations of the new parameters with the standard ones in the posterior distributions. We note that in fig. \ref{triangle_6ug}, $u$ can be found to be slightly positively correlated  to $\omega_{\rm cdm}$, whereas for $n_{\rm s}$, both $u$ and $\Gamma$ happens to be slightly negatively correlated. This is because larger DM-$\nu$ scattering tends to washout DM perturbations at small scales, allowing for more DM energy density and also red-tilt of the primordial power spectrum.

%\vspace{.5cm}
%\noindent\textbf{Role in Hubble tension}

%\noindent 
We now  spend some time on the effect of these new parameters in solving one of the most long-standing tensions between the cosmological and astrophysical observable, the so-called Hubble tension (for history and development in this direction, see for example \cite{1908.03663,2010.04158,2103.01183,2105.05208,2109.01161}
). As is well-known in the community, the values of the Hubble constant $H_0$ as obtained from different observations, do not quite conform with one another. \texttt{Planck 2013} observations found its value to be $H_0=(67.3\pm 1.2)~\rm km ~s^{-1} Mpc^{-1}$ assuming $\Lambda$CDM model \cite{Ade:2013zuv}, whereas on the other hand distance ladder measurement in \texttt{SHOES} project had an estimate of $H_0=(73.8\pm 2.4)~\rm km ~s^{-1} Mpc^{-1}$ \cite{Riess_2011}. Improved analyses in both sides of observations have been unable to narrow down the gap since then, rather widening it. \texttt{Planck 2018} TT,TE,EE+low E+lensing estimate $H_0$ to be $(67.27\pm 0.60)~\rm km ~s^{-1} Mpc^{-1}$ for vanilla $\Lambda$CDM \cite{Aghanim:2018eyx}, \texttt{HST} observation giving $H_0=(74.03\pm 1.42)~\rm km ~s^{-1} Mpc^{-1}$ \cite{Riess:2019cxk}, growing the discrepancy to $4.4-\sigma$.  
Coupled with  R18,  H0LiCOW XIII lensed quasar datasets further increase the tension to $5.3-\sigma$ \cite{Wong:2019kwg}. There has been extensions to $\Lambda$CDM model trying to solve this tension in the literature (see for example \cite{Das:2020wfe}).

In our analysis with the previously mentioned dataset under consideration, we find the mean + ($1-\sigma$) value for $H_0$ is $67.95_{-0.66}^{+0.62}
~\rm km ~s^{-1} Mpc^{-1}$. Therefore, opening up the parameter spaces 
via DM-$\nu$ scattering and DM annihilation helps us in narrowing down the tension slightly.  
%We see that the parameter $\Gamma$ allows for a slightly higher value of $H_0$ for $\Gamma > 0$, due to its slight positive correlation with $H_0$.
This result, although not significant enough to resolve the $H_0$ problem, we note that
the presence of additional relativistic degrees of freedom namely $\nu_s$ and $\phi$ can potentially allow for much higher values of $H_0$, and hence can be a potentially interesting scenario to explore further.

\subsection{6+3 parameter extension and effects of $N_{\rm eff}$}

So far we have been examining a  6+2 parameters model in the light of CMB observations
using \texttt{Planck 2018} dataset (high-l TT+TE+EE, low-l TT, low-l EE).
In order to investigate for the effects of 
DM-$\nu$ scattering and DM annihilation on the effective neutrino species $N_{\rm eff}$, let us extend our 6+2 parameters studies
 to a 6+3 parameter analysis.
 In this case the set of cosmological parameters under consideration are:
$\lbrace \omega_{\rm b},~\omega_{\rm cdm},~\theta_{\rm s},~A_{\rm s},~n_{\rm s},~\tau_{\rm reio},~u,~\Gamma,~N_{\rm eff} \rbrace$.
As in the previous case,  we make use of the modified version of \texttt{CLASS} and the MCMC code \texttt{MontePython} to constrain the parameter spaces as given below.

\begin{figure}
    \centering
    \includegraphics[height=13cm,width=10cm]{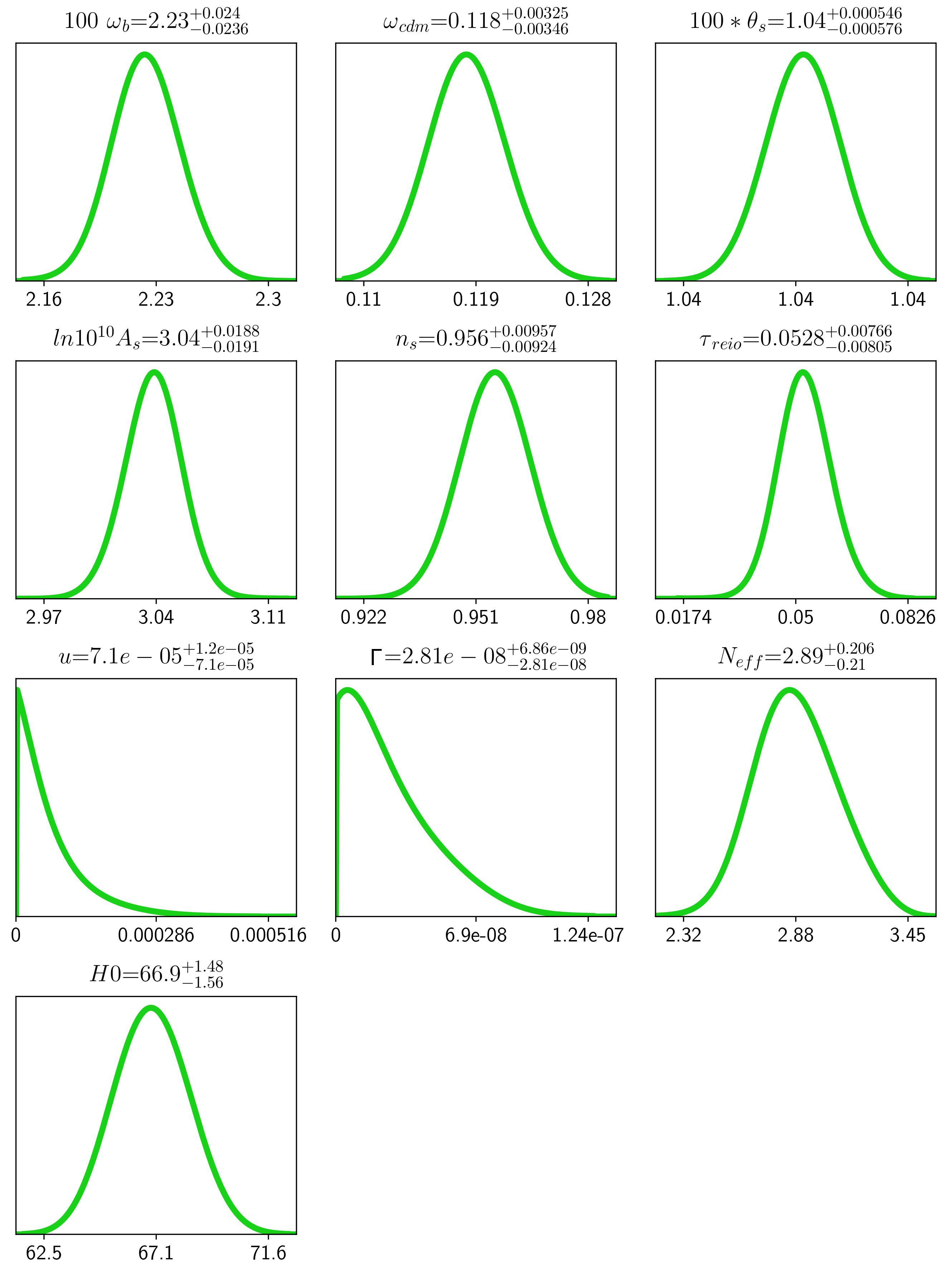}
    \caption{\it 1-d posterior distributions of 6+3 parameter model with parameters $\lbrace
    \omega_{\rm b},~\omega_{\rm cdm},~\theta_{\rm s},~A_{\rm s},~n_{\rm s},~\tau_{\rm reio}, ~u, ~\Gamma ,~N_{\rm eff}\rbrace$ using \texttt{Planck 2018} dataset (high-l TT+TE+EE, low-l TT, low-l EE).}
    \label{1d_6ugn}
\end{figure}

\begin{figure}
    \centering
    \includegraphics[width=1.\textwidth]{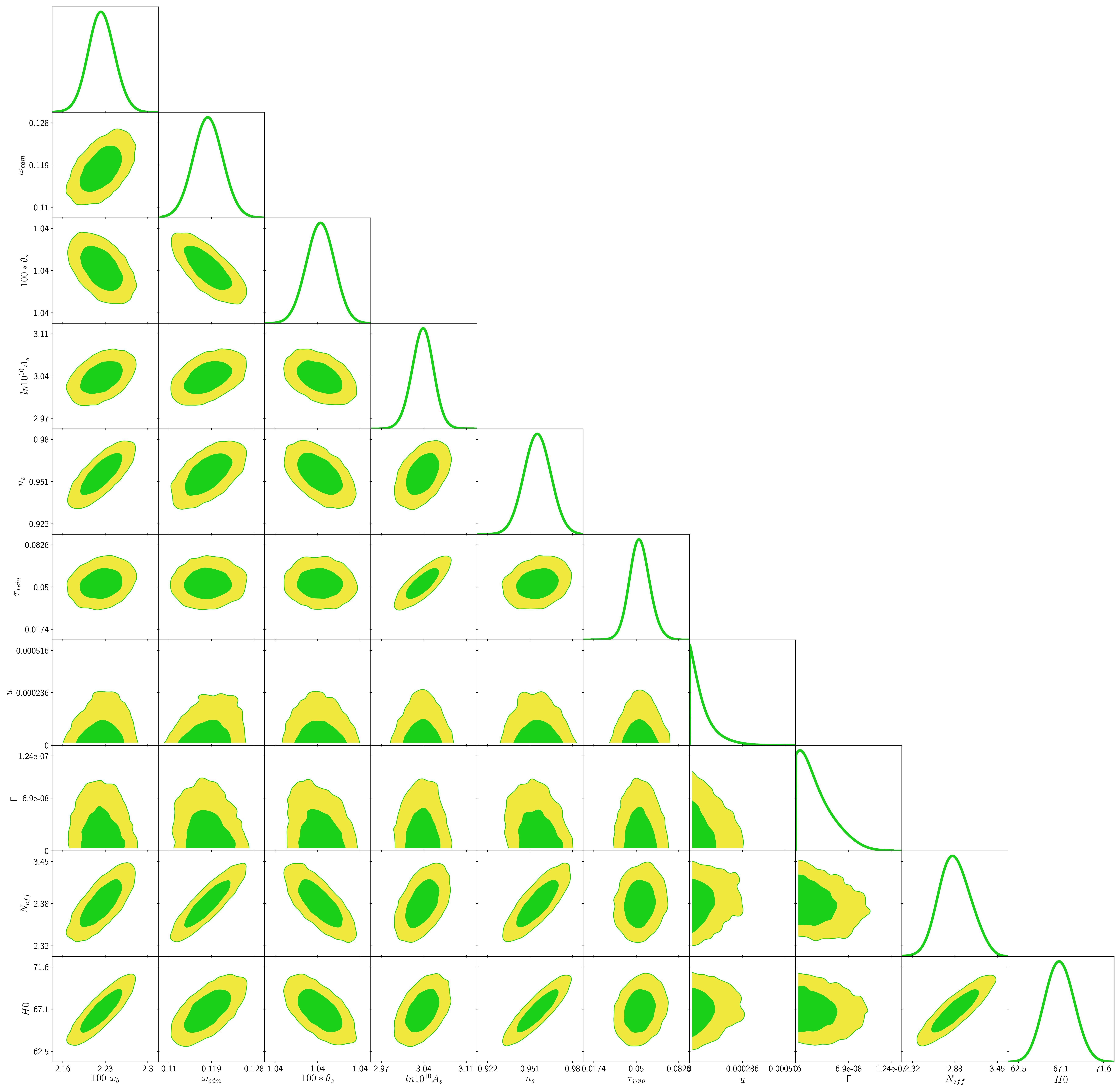}
    \caption{\it 2-d posterior distributions of 6+3 parameter model with parameters $\lbrace \omega_{\rm b},~\omega_{\rm cdm},~\theta_{\rm s},~A_{\rm s},~n_{\rm s},~\tau_{\rm reio}, ~u, ~\Gamma ,~N_{\rm eff}\rbrace$ using \texttt{Planck 2018} dataset (high-l TT+TE+EE, low-l TT, low-l EE).}
    \label{triangle_6ugn}
\end{figure}

The 1-d and 2-d posterior distribution  plots for the 6+3 parameter study are given by figs. \ref{1d_6ugn} and \ref{triangle_6ugn} respectively. 
Statistical results of the nine parameters are given in Table \ref{tab:table2}.
\begin{table}[H] 
\begin{tabular}{|l|c|c|c|c|} 
	\hline 
	Parameter & best-fit & mean$\pm\sigma$ & 95\% lower & 95\% upper \\ \hline 
	$100~\omega{}_{b }$ &$2.24$ & $2.225_{-0.024}^{+0.024}$ & $2.179$ & $2.271$ \\ 
	$\omega{}_{cdm }$ &$0.1189$ & $0.1181_{-0.0035}^{+0.0033}$ & $0.1116$ & $0.1247$ \\ 
	$100*\theta{}_{s }$ &$1.042$ & $1.042_{-0.00058}^{+0.00055}$ & $1.041$ & $1.043$ \\ 
	$ln10^{10}A_{s }$ &$3.04$ & $3.037_{-0.019}^{+0.019}$ & $2.999$ & $3.075$ \\ 
	$n_{s }$ &$0.9633$ & $0.956_{-0.0092}^{+0.0096}$ & $0.9374$ & $0.9742$ \\ 
	$\tau{}_{reio }$ &$0.05145$ & $0.05275_{-0.0081}^{+0.0077}$ & $0.03678$ & $0.06908$ \\ 
	$u$ &$-$ & $8.296\times10^{-5}~(\rm 1-\sigma~upper)$ & $-$ & $0.0002123$ \\ 
	$\Gamma$ &$-$ & $3.502\times10^{-8}~(\rm 1-\sigma~upper)$ & $-$ & $7.192\times10^{-8}$ \\ 
	$N_{eff }$ &$3.001$ & $2.888_{-0.21}^{+0.21}$ & $2.484$ & $3.303$ \\ 
	$H0$ &$67.98$ & $66.88_{-1.6}^{+1.5}$ & $63.86$ & $69.9$ \\ 
	\hline 
\end{tabular}\\\caption{\it Statistical results of 6+3 parameter model with parameters $\lbrace \omega_{\rm b},~\omega_{\rm cdm},~\theta_{\rm s},~A_{\rm s},~n_{\rm s},~\tau_{\rm reio}, ~u, ~\Gamma ,~N_{\rm eff}\rbrace$ using \texttt{Planck 2018} dataset (high-l TT+TE+EE, low-l TT, low-l EE).}
\label{tab:table2} 
\end{table}

%THIS IS THE TABLE WITH MEANS
%\begin{table}[H] \label{table2}
%\begin{tabular}{|l|c|c|c|c|} 
% \hline 
%Parameter & best-fit & mean$\pm\sigma$ & 95\% lower & 95\% upper \\ \hline 
%$100~\omega_{b }$ &$2.254$ & $2.243_{-0.031}^{+0.028}$ & $2.185$ & $2.303$ \\ 
%$\omega_{cdm }$ &$0.1171$ & $0.1187_{-0.0049}^{+0.0046}$ & $0.1094$ & $0.1283$ \\ 
%$100*\theta_{s }$ &$1.042$ & $1.042_{-0.00071}^{+0.00071}$ & $1.04$ & $1.043$ \\ 
%$ln10^{10}A_{s }$ &$3.213$ & $3.181_{-0.057}^{+0.063}$ & $3.058$ & $3.302$ \\ 
%$n_{s }$ &$0.9693$ & $0.9657_{-0.011}^{+0.011}$ & $0.9435$ & $0.9882$ \\ 
%$\tau_{reio }$ &$0.1428$ & $0.1243_{-0.029}^{+0.03}$ & $0.06379$ & $0.1837$ \\ 
%$u$ &$1.598e-05$ & $0.0001135_{-0.00011}^{+2.2e-05}$ & $3.902e-09$ & $0.0003351$ \\ 
%$\Gamma$ &$3.886e-08$ & $6.181e-08_{-6.2e-08}^{+1.7e-08}$ & $1.759e-11$ & $1.592e-07$ \\ 
%$N_{eff}$ &$3.041$ & $3.058_{-0.29}^{+0.28}$ & $2.493$ & $3.62$ \\ 
%$H0$ &$69.3$ & $69_{-1.9}^{+1.7}$ & $65.5$ & $72.65$ \\ 
%\hline 
% \end{tabular} \\\caption{\it Statistical results of 6+3 parameter model with parameters $\lbrace \omega_b,~\omega_{cdm},~\theta_s,~A_s,~n_s,~\tau, ~u, ~\Gamma ,~N_{eff}\rbrace$ using \texttt{Planck 2018 high-l TT, BOSS-BAO 2014} data-sets.}
%%$-\ln{\cal L}_\mathrm{min} =379.81$, minimum $\chi^2=759.6$ \\ 
%\end{table}

Similar to the 6+2 parameter analysis, in the 6+3 parameter case, we obtain  $H_0= 66.88_{-1.6}^{+1.5}
~\rm km~s^{-1}~Mpc^{-1}$. We observe that, the $1-\sigma$ width of the $H_0$ posterior distribution is higher than that obtained in 6+2 parameter scenario. This is mostly because of the inclusion of one more parameter $N_{\rm eff}$. This possibility of extra relativistic species is natural in the analysis due to presence of additional particles, namely $\nu_s$ and $\phi$ in the particle spectrum. It should also be noted that  higher $N_{\rm eff}$ allows for higher $H_0$ value, due to the strong positive correlation between $H_0$ and $N_{\rm eff}$, as is well known from other literature \cite{Bernal:2016gxb,DiValentino:2021izs,Vagnozzi:2019ezj}.
 
%%%%%%%%%%%%%%%%%%%%%%%%%%%%%%%%%%%%%%%%%%%%%%%%%%%%%%%%%%%%

\section{Viable DM model}
\label{model}

In order to understand how the bounds on the cosmological parameters obtained from our analysis  is helpful in constraining  particular DM models, we discuss a typical model involving DM \& sterile neutrinos. We particularly concentrate on light ($\sim$ eV) sterile neutrinos since they are motivated from neutrino Small Base Line (SBL) anomalies.  
Note that in SBL experiments, \texttt{LSND} \cite{Athanassopoulos:1995iw,Aguilar:2001ty} and \texttt{MiniBooNE} \cite{Aguilar-Arevalo:2018gpe} observed excess in $\bar{\nu}_{\mu}\rightarrow \bar{\nu}_e$ channel, \texttt{MiniBooNE} have also indicated an excess of $\nu_e$ in the $\nu_{\mu}$ beam. Within a 3+1 framework, these results hints towards the existence of a sterile neutrino with eV mass. Recently \texttt{MiniBooNE} excess has reached 4.8-$\sigma$ with more data available till date \cite{Aguilar-Arevalo:2018gpe}.

Although debatable in 3+1 framework,  such a light additional sterile neutrino, with mixing 
$\sin (\theta_{\rm m}) \lesssim \mathcal{O}(0.1)$ with the active neutrino species, is 
consistent with constraints from various terrestrial neutrino experiments. But, presence of an additional sterile neutrino species, if thermalised in the early Universe, is tightly constrained from $N_{\rm eff}$ bound of Big Bang Nucleosynthesis (BBN). However, this bound can be respected in presence of additional interactions among the sterile neutrinos, dubbed as "secret"\footnote{In our scenario, "secret" interactions are mediated by light pseudoscalar particles.} interactions \cite{Hannestad:2013ana,Archidiacono:2014nda,Archidiacono:2015oma, Archidiacono:2016kkh, Chu_2015}. These interactions mediated by light bosonic mediators suppress the sterile neutrino production in early Universe upto BBN via Mikheyev-Smirnov-Wolfenstein (MSW) like effect, hence evading the $N_{\rm eff}$ bound.

\subsection{A model motivated by particle physics experiments}

 In one of the earlier works by the same authors \cite{Paul_2019},  primordial abundance of such light dark species as mentioned above, has been investigated in details considering particle production from inflationary (p)-reheating.  %and identified the parameter space compatible with such a set-up has been studied extensively. 
In the same vein,  we will like  to consider a somewhat similar scenario where the interaction terms in eq.(\ref{eq:lag}) are present, leading to DM-$\nu$ scattering \& DM annihilation which we constrain by CMB analysis. This will translate the model-independent bounds on $u$ and $\Gamma$ obtained in section \ref{analysis} to the parameter space of the model. %, namely, in terms of DM mass $M_{\Psi}$ \& DM-pseudoscalar coupling $g_{\Psi}$. 
In the rest of the paper, we will like to engage ourselves in analysing and constraining a particular DM model and comment on its viability in the light of recent observations based on our model-independent analysis done so far.

The model under our consideration is a simple extension of a model used to accommodate light eV scale sterile neutrinos (required to explain the neutrino anomalies) in cosmology. For this model to be compatible with cosmology, an extra pseudo-scalar particle $\phi$ is introduced \cite{Hannestad:2013ana,Archidiacono:2014nda,Archidiacono:2015oma, Archidiacono:2016kkh}. This extra interaction causes a MSW type effect and suppresses the production of $\nu_s$ through oscillation until BBN, hence evading the $N_{\rm eff}$ bound of the same. In the current article we introduce a Dirac fermion $\Psi$ which would be used as the DM in this article. The Lagrangian is given by,
\be\label{eq:lag}
- \mathcal{L} \supset g_s \bar{\nu}_s \gamma_5 \nus \phi+ g_{\Psi} \bar{\Psi}\gamma_5 \Psi \phi
\ee

The parameter $g_s$ is required to be $\gtrsim\mathcal{O}(10^{-5})$ for the suppression of sterile neutrinos until BBN \cite{Archidiacono:2016kkh}, similar to the above mentioned articles.

%CDM particles with sub-MeV mass are difficult to produce in standard thermal production mechanisms, as they are relativistic during BBN and contribute to $N_{\rm eff}$, which is strictly constrained. However, a thermal production mechanism can be engineered for this sub-MeV scale DM \cite{Berlin_2018,Berlin_2019}, if the thermalisation happens after BBN, hence evading the BBN constraints.

We note that, the smallest observable mode ($l=2500$) in CMB enters the Hubble horizon when the temperature of Universe was $\mathcal{O}(10~\rm eV)$, so the perturbation equations which we introduced in section \ref{sec_pert}, are applicable for all models where the CDM particles have been produced and are non-relativistic before that temperature.

%Although we are mainly interested in thermally produced sub-MeV DM particles in this article, it is worth mentioning that the treatment is also applicable for any model where there is DM-$\nu$ scattering and an Sommerfeld enhanced annihilation of DM particles  into any invisible radiation like species. So, we can use the results to constrain models where the DM has been produced non-thermally, for example during (p)reheating. We should mention that, although the results may be cast in an model-independent manner, the constraints in this model may not be applicable for condensate-like DM or non-cold DM where the evolution of the perturbations follow different evolution equations.

\begin{figure}
	\centering
	\includegraphics[height=4cm,width=6cm]{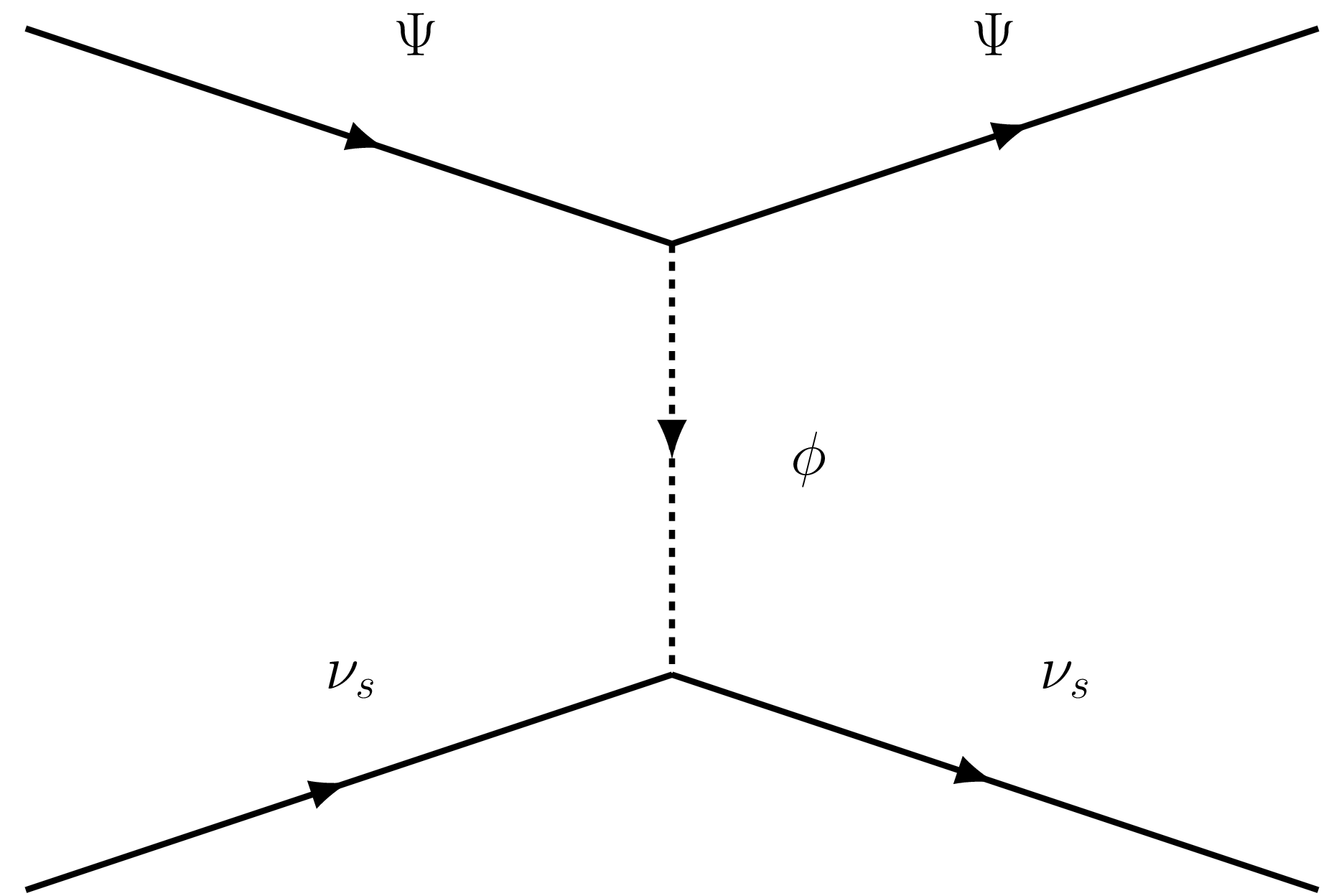}
	\includegraphics[height=4cm,width=6cm]{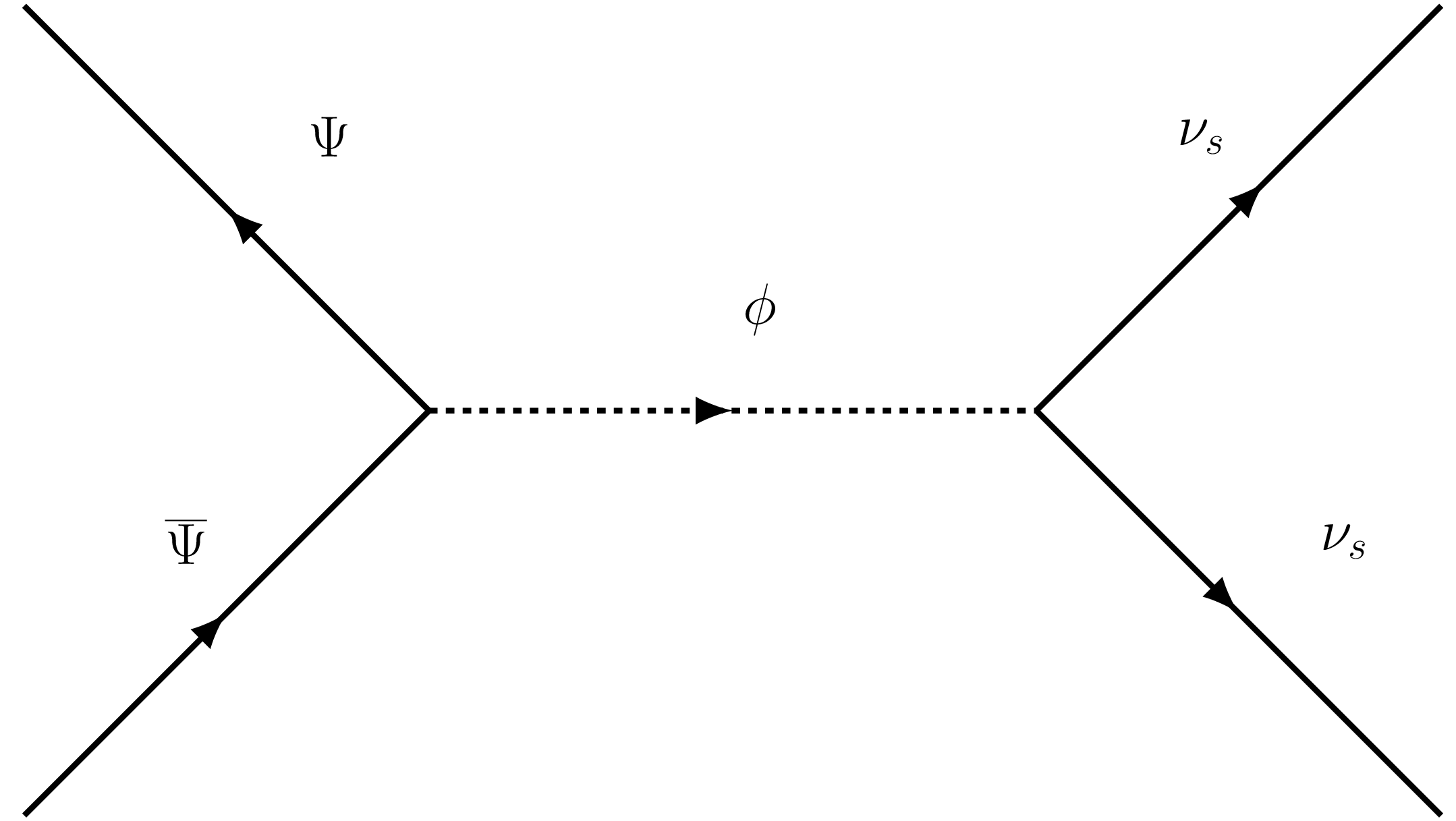}
	\caption{\it Feynman diagrams representing DM--$\nu_s$ scattering and DM annihilation into $\nu_s$.}
	\label{fig:feyn}
\end{figure}

The model gives rise to DM-$\nu$ scattering through the pseudo-scalar mediator and the annihilation of DM particles into $\nu_s$ or $\phi$ through $s$ or $t$ channel diagram with $\phi$ or DM mediator. In fig. \ref{fig:feyn}, we show the Feynman diagrams corresponding the DM-$\nu_s$ scattering and DM annihilation into $\nu_s$ according to our model. As we mentioned before, our aim in this section is to constrain the model from effect of these kinds of interactions in cosmological observables.
This we will do in the subsection that follows.

\subsection{Constraining model parameters}
%\label{}

\begin{figure}[h!]
	\centering
	\includegraphics[width=0.8\linewidth]{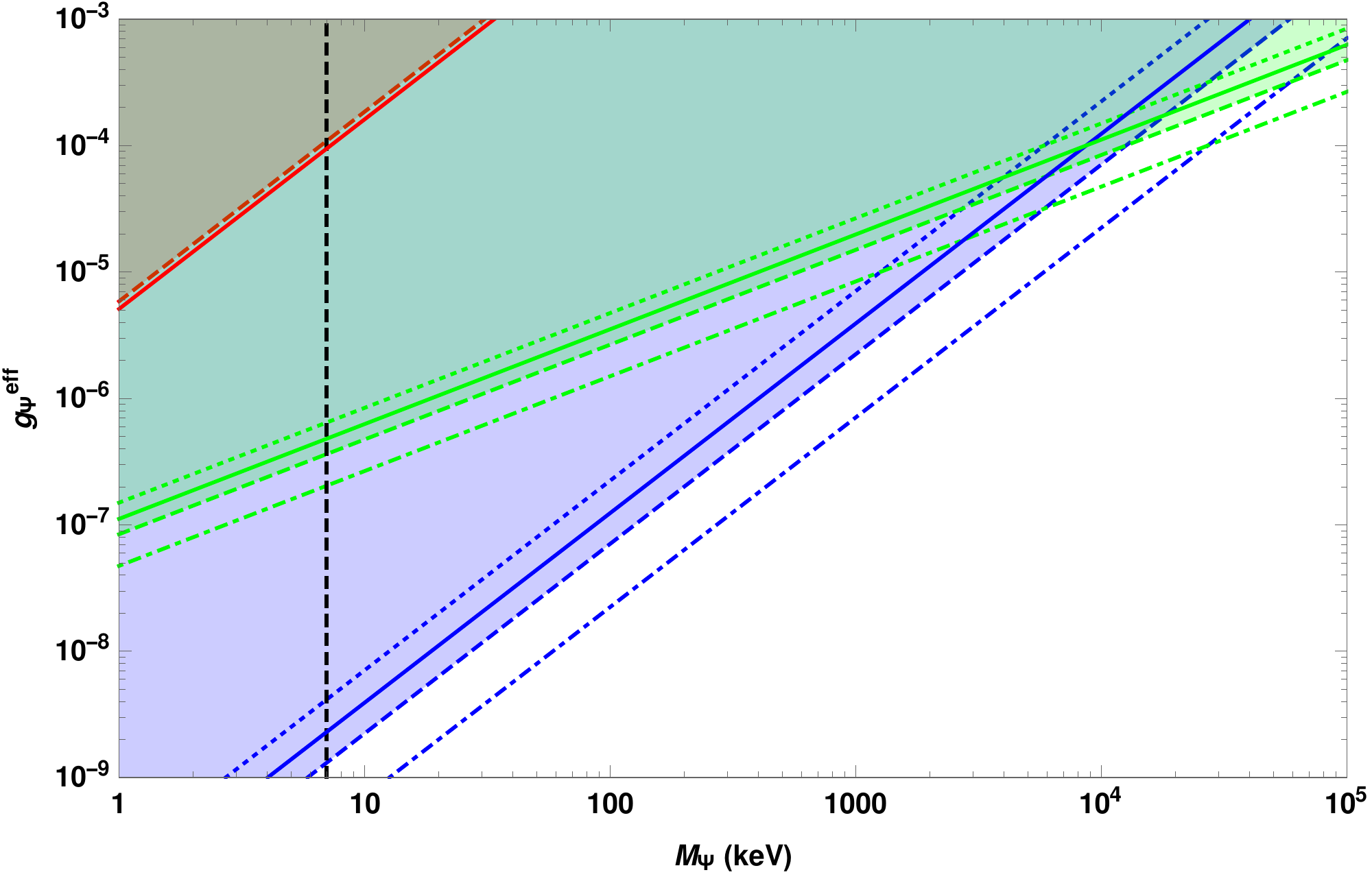}
	\caption{\it Constraints on $g_{\Psi}^{\rm eff}$ vs $M_{\Psi}$ plane from DM-$\nu$ scattering and DM annihilation. We have used $g_{\Psi}^{\rm eff}=g_{\Psi}\sin^2(\theta_{\rm m})$ for the scattering process and  $g_{\Psi}^{\rm eff}=g_{\Psi}$ for the annihilation processes to keep them on the same footing, as during scattering the DM particles scatter to the active neutrinos through mixing with $\nu_s$, whereas during annihilation, the more stringent bound comes from annihilation of DM into $\nu_s$. The red, blue and green solid lines correspond to bounds from scattering and annihilation of DM into sterile neutrinos and pseudoscalars respectively in the scenario without Sommerfeld enhancement. The dashed red line and dotted, dashed, dot-dashed blue and green lines correspond to bounds from scattering and annihilation of DM into sterile neutrinos and pseudoscalars for $\langle S \rangle=100,~1000,~10000$ respectively at $z=1100$. We have used $g_s=10^{-4}$ and $\theta_{\rm m}=0.1$ for these plots. The vertical dashed line at $M_\Psi\sim 7$ keV corresponds to the lower mass bound of fermionic DM from Lyman-$\alpha$ observations \cite{Garzilli:2018jqh}.}
    \label{fig:scatter_anni3}
\end{figure}

The 1-d posterior bounds on $u$ from the cosmological observations can be translated to parameter space of the particle model through eq.(\ref{u_nudm}). In the current model, $$\sigma_{\Psi-\nu}=\frac{g_s^2 g_{\Psi}^2}{64\pi M_{\Psi}^2}\sin^4(\theta_{\rm m}).$$ 
Here, the mixing term between active and sterile neutrinos $\sin(\theta_{\rm m})$ is present because the pseudo-scalar is coupled directly to the sterile neutrino according to the Lagrangian (\ref{eq:lag}), and hence coupled to the active neutrinos through mixing. So, in order to describe scattering rate of DM to active neutrinos, which is indeed the case of perturbation evolutions in section \ref{sec_pert}, we require to take the mixing angle into account. In order to keep the bound from scattering on the same footing with that of annihilation, we define $g_{\Psi}^{\rm eff}=g_{\Psi}\sin^2(\theta_{\rm m})$ to plot the constraints from scattering in fig. \ref{fig:scatter_anni3}. We also mention that, although the "secret" interaction term with the pseudoscalar suppresses the mixing angle until BBN, for our study we can assume this mixing angle to be sizable ($\theta_{\rm m}\sim0.1$), as the modes we are interested in, enters the Hubble horizon far after BBN.

Similarly, constraints on $\Gamma$ translates to the $g_{\Psi}$ vs $M_{\Psi}$ parameter space, given the Sommerfeld enhanced thermally averaged annihilation cross-section $\langle\sigma v\rangle$ \cite{Tulin:2013teo,Cassel:2009wt,Aarssen:2012fx} of DM into sterile neutrinos or pseudoscalar particles. We  assume that DM freezes out at a temperature $\sim \frac{M_\Psi}{10}$, and follows Maxwell-Boltzmann distribution with temperature $T_\Psi$ evolving as $\sim \frac{1}{a^2}$ since then. The tree level thermally averaged annihilation cross-section of DM into sterile neutrinos or pseudoscalar particles respectively are given by $\langle\sigma v\rangle_{\rm tree}=\frac{g_\Psi^2 g_s^2}{64 M_\Psi^2}~\rm{or}$ $\frac{g_\Psi^4}{64 M_\Psi^2}$, assuming their masses to be negligible with respect to DM. It is worth mentioning that, cosmologically we do not distinguish between neutrino and pseudoscalar species, both of them behaving as relativistic species and contributing to $N_{\rm eff}$. However, when considering the annihilation of DM into different species, we get different bounds due to difference in expressions of annihilation cross-sections. The Sommerfeld enhanced thermally averaged annihilation cross-section is then given by,
$$\langle\sigma v\rangle=\langle S \sigma v\rangle_{\rm tree}=\langle S\rangle \langle\sigma v\rangle_{\rm tree},$$
where $\langle S\rangle$ is the thermally averaged Sommerfeld enhancement factor. At some specific epoch, this factor $\langle S\rangle$ can be chosen as a free parameter if we assume an additional mediator with sizable coupling strength to DM. However if there is no such additional species in the model, $\langle S\rangle$ boils down to $\langle S\rangle=\langle\frac{g_\Psi^2}{4 v}\rangle=\frac{g_\Psi^2}{\sqrt{8\pi}}\sqrt{\frac{M_\Psi}{T_\Psi}}$. We have checked that, this expression does not lead to enhancement due to the factor $g_\Psi^2$, which becomes small when we use the constraints of $\Gamma$. For that reason, we keep $\langle S\rangle$ as a free parameter. %$\langle\sigma v\rangle_{\rm tree}=\frac{g_\Psi^2 g_s^2}{64 M_\Psi^2}~\rm{or}$ $\frac{g_\Psi^4}{64 M_\Psi^2}$ is the tree level thermally averaged annihilation cross-section of DM into sterile neutrinos or pseudoscalar particles respectively, assuming their masses to be negligible with respect to DM. For the DM temperature $T_\Psi$ in the expression of $\langle S\rangle$, we assume the DM to freeze-out at a temperature of $\sim \frac{M_\Psi}{10}$, and evolve as $\sim \frac{1}{a^2}$ since then.

\begin{figure}[h!]
	\centering
	\includegraphics[width=0.8\linewidth]{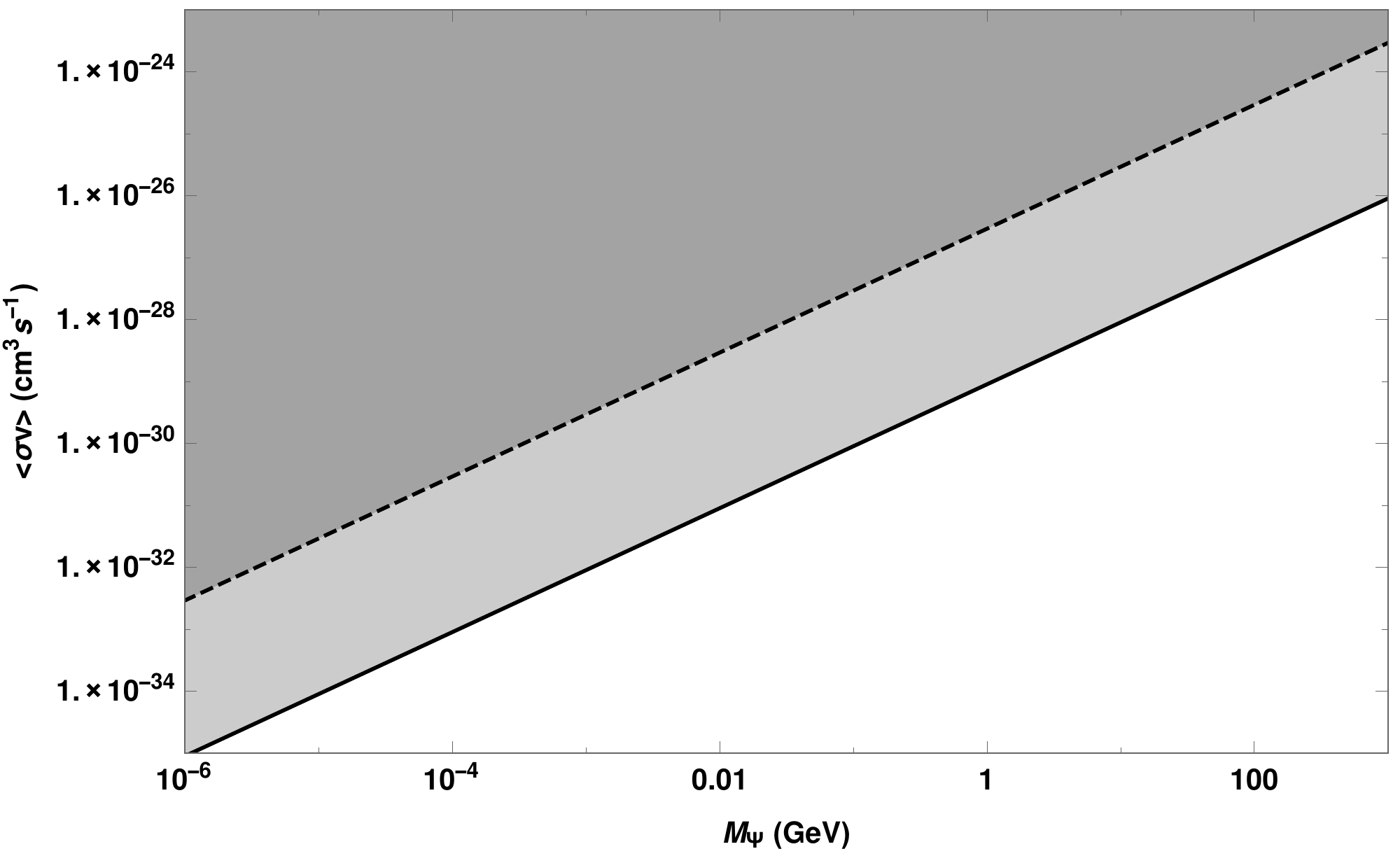}
	\caption{\it Constraints on $\langle \sigma v\rangle$ vs $M_{\Psi}$ plane. The solid and dashed lines correspond to annihilation of DM particles into invisible species for a scenario without Sommerfeld enhancement and with Sommerfeld enhancement  at $z=1100$ respectively.
}
\label{fig:sigmav_vs_Mpsi}
\end{figure}

In presence of both scattering and annihilation in our model,  we are able to map the $1-\sigma$ bounds of posterior distributions of $\Gamma$\footnote{The 1-$\sigma$ upper bound of $\Gamma=3.204\times10^{-8}$ translates to a value of $\langle\sigma v\rangle=2.91\times 10^{-25}\rm~cm^3~s^{-1}$ at last scattering surface for $M_\Psi=100~\rm GeV$, which is comparable to the upper bound of thermally averaged DM annihilation cross-section to invisible channel \cite{Cui:2018imi} and visible particles that contribute to the heating of photon bath \cite{Aghanim:2018eyx}. In comparison, for the scenario when there is no enhancement in the DM annihilation rate, 1-$\sigma$ upper bound of $\Gamma=8.96\times10^{-14}$ translates to a value of $\langle\sigma v\rangle=8.96\times 10^{-28}\rm~cm^3~s^{-1}$. Bounds of $\langle\sigma v\rangle$ as a function of DM mass $M_{\Psi}$ is shown in fig. \ref{fig:sigmav_vs_Mpsi}.} and $u$  of fig. \ref{triangle_6ug} into the $g_{\Psi}^{\rm eff}$ vs $M_{\Psi}$ parameter space, using $g_s=10^{-4}$ and $\theta_{\rm m}=0.1$. In fig. \ref{fig:scatter_anni3} we consider the DM-$\nu$ scattering and DM annihilation to both sterile neutrinos and pseudoscalars at one go. Note that, we have used $g_{\Psi}^{\rm eff}=g_{\Psi}\sin^2(\theta_{\rm m})$ for the scattering process and  $g_{\Psi}^{\rm eff}=g_{\Psi}$ for the annihilation processes to keep them on the same footing. The red, blue and green non-solid lines correspond to bounds from scattering and annihilation of DM into sterile neutrinos and pseudoscalars respectively, given different choices of $\langle S\rangle$ values. For comparison, we have also included the solid lines - the bounds for the scenario when there is no enhancement in the DM annihilation rate. Note that for $g_\Psi>g_s$, the DM particles prefer to annihilate into pseudoscalars rather than sterile neutrinos, which is clear from fig. \ref{fig:scatter_anni3}. We note that for our choice of benchmark point, annihilation rate gives more stringent bounds than scattering rate in the $g_{\Psi}$ vs $M_{\Psi}$ parameter space for fermionic DM with mass $M_\Psi> 7$ keV \cite{Garzilli:2018jqh}. However this is a model dependent statement and the result may also change for other benchmark points (for example, higher value of $g_s$).
%However, these bounds from $u$ and $\Gamma$ may become comparable if we choose higher values of $g_s$ and $\theta_{\rm m}$.

\section{Conclusions}
\label{conclusion}

Let us now summerize the salient points of the present analysis.

\begin{itemize}
    \item In this article, for the first time we analyse the effect of DM-$\nu$ scattering and DM annihilation into invisible \textit{radiation-like} species in one go in cosmological observables and constrain the two scenarios using  \texttt{Planck 2018} dataset (high-l TT+TE+EE, low-l TT, low-l EE). To materialize this, we formulate the scenario with perturbations in Newtonian gauge and modify the publicly available code \texttt{CLASS} to incorporate the effects.
We further use the MCMC code \texttt{MontePython} to estimate the posterior distribution of the parameters under consideration.

\item For cosmological aspects, we observe that the new parameters $u$ and $\Gamma$ not present in the vanilla $\rm \Lambda$CDM model, get only upper bounds ($\sigma_{\Psi-\nu}<6.75\times 10^{-29}\rm~cm^2$, $\langle\sigma v\rangle<2.91\times 10^{-25}\rm~cm^3~s^{-1}$ at last scattering surface for $M_\Psi=100~\rm GeV$), the posteriors being also consistent with zero. This reiterates that the present cosmological observables are well explained by the simplest possible model $\rm \Lambda$CDM. However, we notice that, by opening up the parameter space for the data-sets under consideration, both the 6+2 and 6+3 parameter models give the mean + ($1-\sigma$) value for $H_0$ to a slightly higher value than that possible in vanilla $\Lambda$CDM model, thereby narrowing down the $H_0$ estimate between cosmological and astrophysical observations. We also observe that, although these non-standard effects does not significantly help to reconcile the $H_0$ problem in a straight forward manner, the tension can be relaxed to some extent within this model if we open the $N_{\rm eff}$ parameter during Monte-Carlo simulation. 

We mention that the simplistic choice of neutrinos being massless in our analysis forbids us from doing the analysis with $\Sigma m_\nu$ (sum of neutrinos masses) as a free parameter, which may have degeneracies with the other parameters. We look forward to study that case in a future analysis.

\item Further, we study a particular DM model that allows us incorporate eV scale sterile neutrinos, required to solve neutrino anomalies, in cosmology and also gives us a possible mechanism to thermally produce DM particles. This model gives rise to DM-$\nu$ scattering and also may result in Sommerfeld enhanced DM annihilation due to the presence of the light pseudo-scalar particle in the particle spectrum. Model like this where all the new interactions are in the invisible sector are hard to constrain in standard laboratory or astrophysical probes. So, one possible way to constrain them is through cosmological observations. We found that Sommerfeld enhancement of DM annihilation via the same light pseudoscalar particle is not allowed by the datasets we used, however the enhancement factor $\langle S\rangle$ may arise via an additional light particle. We show how one can map  the $1-\sigma$ constraints  on the interaction terms $u$ and $\Gamma$ of fig. \ref{triangle_6ug}. to the 
corresponding constraints on the parameters   $g_{\Psi}^{\rm eff}-M_{\Psi}$ of this particular model in fig. \ref{fig:scatter_anni3}. %\ref{table1}.
%We, constrain the $g_{\Psi}-M_{\Psi}$ %and $\langle\sigma v\rangle _{anni}-M_{\Psi}$ 
%plane in fig:  \ref{fig:scatter_anni2} by mapping 1-$\sigma$ and 2-$\sigma$ bounds on $u$ and $\Gamma$ in table. 1. %\ref{table1}.

\end{itemize}

In a nutshell, the present analysis not only constrains the particle physics motivated DM-neutrino scattering and DM annihilation scenarios in a somewhat generic set-up using latest  cosmological datasets but also  helps us investigate for any possible effects of those phenomena on cosmological parameters and subsequently compare them with vanilla $\Lambda$CDM cosmology.

\section*{Acknowledgement}
Authors gratefully acknowledge the use of publicly available code \texttt{CLASS} and \texttt{MontePython} and thank the computational facilities of Indian Statistical Institute, Kolkata. AP and AG thank Balakrishna S. Haridasu and AG thanks Miguel Escudero %(SISSA, Trieste)
for useful discussions. Authors also thank the anonymous referee for useful suggestions. AP thanks CSIR, India for financial support through Senior Research Fellowship (File no. 09/ 093 (0169)/ 2015 EMR-I). AC acknowledges support from DST, India, under grant number IFA 15 PH-130.  
SP thanks Department of Science and Technology, Govt. of India for partial support through Grant No. NMICPS/006/MD/2020-21.

%\section{References}
\bibliographystyle{utphys}
\bibliography{main_v3}

\end{document}